\documentclass[twocolumn]{aastex631}

\usepackage{amsmath,amsfonts,amssymb}
\usepackage{mathrsfs}  % package for the "curly" fonts
\usepackage{url}
\usepackage{multirow}
\usepackage{graphicx}
\graphicspath{{plots/}}
\usepackage{xcolor}
\usepackage[normalem]{ulem}
\usepackage{fancyhdr}

\usepackage{hyperref}
\hypersetup{
    colorlinks = true,
    linkcolor = {blue},
    citecolor = {blue},
    urlcolor = {blue},
    linkbordercolor = {white},
    }

\begin{document}

\title{Inferring neutron star merger ejecta morphologies with kilonovae}

\author{Brendan L. \surname{King}}
\email{king0690@umn.edu}
\affiliation{School of Physics and Astronomy, University of Minnesota, Minneapolis, MN 55414}\affiliation{Theoretical Division, Los Alamos National Laboratory, Los Alamos, NM 87545, USA}

\author{Soumi \surname{De}}
\affiliation{Theoretical Division, Los Alamos National Laboratory, Los Alamos, NM 87545, USA}

\author{Oleg \surname{Korobkin}}
\affiliation{Theoretical Division, Los Alamos National Laboratory, Los Alamos, NM 87545, USA}

\author{Michael W. \surname{Coughlin}}
\affiliation{School of Physics and Astronomy, University of Minnesota, Minneapolis, MN 55414}

\author{Peter T. H. \surname{Pang}}
\affiliation{Nikhef, Science Park 105, 1098 XG Amsterdam, The Netherlands}
\affiliation{Institute for Gravitational and Subatomic Physics (GRASP), Utrecht University, Princetonplein 1, 3584 CC Utrecht, The Netherlands}

\date{\today}

\begin{abstract}
In this study we incorporate a new grid of kilonova simulations produced by the Monte Carlo radiative transfer code SuperNu in an inference pipeline for astrophysical transients, and evaluate their performance.
These simulations contain four different two-component ejecta morphology classes.
We analyze follow-up observational strategies by Vera Rubin Observatory in optical, and James Webb Space Telescope (JWST) in mid-infrared (MIR).
Our analysis suggests that, within these strategies, it is possible to discriminate between different morphologies only when late-time JWST observations in MIR are available.
We conclude that follow-ups by the new Vera Rubin Observatory alone are not sufficient to determine ejecta morphology.
Additionally, we make comparisons between surrogate models based on radiative transfer simulation grids by SuperNu and POSSIS, by analyzing the historic kilonova AT2017gfo that accompanied the gravitational wave event GW170817.
We show that both SuperNu and POSSIS models provide similar fits to photometric observations.
Our results show a slight preference for SuperNu models, since the wind ejecta parameters recovered with these models are in better agreement with expectations from numerical simulations. 
\end{abstract}

%%%%%%%%%%%%%%%%%%%%%%%%%%%%%%%%%%%%%%%%%%%%%%%%%%%%%%%%%%%%%%%%%%%%%%%%%%%%%%
\section{Introduction}
\label{sec:intro}

Neutron star mergers (NSMs) are one of the most energetic events in the universe hosting a plethora of phenomena that provide deep insights in the physics of matter at supra-nuclear densities, gravity in the strong-field regime, and heavy element nucleosynthesis via the rapid neutron capture process ($r$-process) \citep{Burns_2020}.
The $r$-process is a complex network of nuclear reactions responsible for the creation of approximately half of the atomic nuclei heavier than iron, and NSMs are one of the leading candidates for its astrophysical site, offering conditions of extreme neutron richness and rapid decompression, perfect for the $r$-process~\citep{freiburghaus99}.
Many big questions about the origin of the $r$-process remain unanswered or require clarification.
Can $r$-process nucleosynthesis from NSMs alone account for the observed $r$-process residuals in the Solar system~\citep{cowan21}?
What are the macroscopic properties---nuclear heating rates, compositions, masses, morphologies, and expansion velocities---of individual ejection channels in NSMs?
How much can be learned about these macroscopic properties from the analysis of a kilonova---an electromagentic signal in optical and infrared, emitted by the radioactive decay of $r$-process elements~\citep{metzger17y}?
In this study, we address the last question by investigating how much information about ejecta morphology can be extracted from photometric observations of a kilonova.

There are multiple channels in which neutron-rich material can be expelled from the system, power a kilonova, and contribute to the Galactic nucleosynthesis.
Here we will group them in two components: the \emph{dynamical ejecta} and the \emph{wind}, both of which fuel the kilonova.
As suggested by numerical relativity simulations, the first component is likely composed of neutron-rich material that is ripped off neutron stars by tidal forces~\citep{rosswog13,Dietrich_2017,Hotokezaka_2015}, while the second component is produced mainly after the merger, by the hypermassive remnant or its accretion disk~\citep{metzger17y,fernandez15,miller19b}.
Due to its neutron richness, the dynamical ejecta component is lanthanide-rich, while the wind is lanthanide-free.
The extreme bound-bound opacities of lanthanides render the kilonova from dynamical ejecta component slower and redder, peaking in the near-infrared bands, while the wind component is blue and peaks on a timescale of a day~\citep{metzger10,grossman14}.
This two-component picture is a convenient proxy for kilonova models.
Although it does not separately include all possible ejection channels, it accounts for the essential physical characteristic---lanthanide content---which subsequently allows us to tie observations to nuclear physics.
Accurately inferring the properties of these two components from kilonova observations is an important step towards determining the total nucleosynthesis in the ejecta.

To date, there has been one definitive detection of a kilonova dubbed AT2017gfo with a coincident gravitational wave (GW) signal named GW170817~\citep{abbott17a, coulter17, arcavi17n, lipunov17o, soaressantos17, tanaka17, tanvir17, valenti17, evans17, alexander17, margutti17, Sugita_2018, troja17, chornock17, Andreoni_2017, hul17, cowperthwaite17, diaz17, drout17, kasliwal17a, kasliwal17b, pian17, shappee17, smartt17, Utsumi_2017}.
Following this one confirmed detection of a kilonova, concerted effort has been made to quickly capture data for a subsequent event.
Wide-field surveys like the Zwicky Transient Facility are the first choice for finding a kilonova after a GW trigger.
These surveys are done with telescopes with wide fields-of-view (FOVs) to scan a probable region of the sky promptly following a high-likelihood event from a GW detector.
Large localization areas from GW triggers require a wide FOV at the expense of the limiting detectable magnitudes, allowing for kilonova discovery only in our nearest neighborhood ($\le$100-300 Mpc~\citep{coughlin19,anand20,Kiendrebeogo_2023}).
It also means spectroscopic follow-up is rarely possible in a timely fashion and thus data from these surveys usually take the form of broadband magnitudes for the set of filters on the instrument.

Accurate parameter estimation of kilonovae is important to extract optimal information from the upcoming observations in the era of multi-messenger astronomy.
The inverse problem---inferring the physical properties of the merging system from observed data---poses a formidable challenge. 
Statistical methods, particularly Bayesian inference and machine learning (ML), have emerged as powerful tools to tackle this complexity~\citep{radice19,pang22}.

The Nuclear Multi-Messenger Astrophysics (NMMA) Python framework has been developed as a pivotal computational tool for researchers in the field of astrophysics and nuclear physics.
Designed to integrate a broad range of statistical methods, NMMA serves as a comprehensive platform for parameter estimation, data assimilation, and uncertainty quantification \citep{pang22}.
The framework incorporates Bayesian inference algorithms, ML techniques such as neural networks, and support vector machines (SVMs).
Recent advances have expanded its capabilities to include real-time data processing, thereby enhancing its utility as a multi-messenger analysis tool~\citep{barna2024}.
NMMA has been used to place constraints on the local Hubble constant~\citep{Kiendrebeogo_2023} and has also been used to obtain the tightest constraints on the radius of a 1.4~M$_\sun$ neutron star, $R_{1.4}$, which is closely related to the nuclear equation of state \citep{Dietrich_2020,pang22}.

In this paper, we leverage the capabilities of the NMMA framework to perform kilonova ``tomography'', namely inferring morphology of the two-component ejecta, based on photometric observations.
We use state-of-the-art kilonova models computed with advanced radiative transport codes POSSIS~\citep{bulla19} and SuperNu~\citep{wollaeger13,wollaeger14}, which employ morphologies informed by NSM and accretion disk simulations~\citep{Dietrich_2017,miller19a}.
Kilonova models created with SuperNu offer a more sophisticated representation of underlying physics, including detailed nuclear heating, thermalization, nucleosynthetic yields, and elemental opacities~\citep{wollaeger19,fontes20}.
Our main objective is to critically assess the discriminating power of these models when integrated into the NMMA framework, with the aim of providing a more robust understanding of the origin of the different kinds of the $r$-process, in our case represented by two different components.
In particular, the so-called ``strong'' $r$-process, most likely synthesized in the tidal ejecta where neutron-rich material from the upper layers of the star undergoes cold decompression, exhibits a remarkably consistent nucleosynthesis pattern~\citep{korobkin12}, previously connected to the robust pattern of $r$-process in metal-poor $r$-process--enriched halo stars \citep{cowan21}.
As the current theoretical models suggest that the strong and weak $r$-processes are produced in two different components, our study addresses the question of whether and to what extent it is possible to distinguish different morphologies in the observed kilonova photometry.

This paper is organized as follows:
in Section~\ref{subsec:validation} we present our chi-squared and parameter recovery analyses to determine the validity of our trained neural networks.
A secondary objective is to evaluate the predictive power of these models when applied to simulated data sets matching next generation transient follow-up strategies.
In Section~\ref{subsec:morph} we look at the upcoming strategy to be used by the Vera Rubin Observatory when doing kilonova follow-up and assess whether photometric observations from Vera Rubin alone can effectively discern parameters of the two-component kilonova models.
We then add simulated photometric observations with the new James Webb Space Telescope (JWST) and test how the addition of the mid-infrared points can aid in identifying kilonova morphology.
Finally, in Section~\ref{subsec:at2017gfo} we apply the new SuperNu two-component model grid to analyze the historic event AT2017gfo and compare to an analysis done with the POSSIS models.

%%%%%%%%%%%%%%%%%%%%%%%%%%%%%%%%%%%%%%%%%%%%%%%%%%%%%%%%%%%%%%%%%%%%%%%%%%%%%%
\section{Methods}
\label{sec:methods}
We use the state-of-the-art multi-messenger pipeline, Nuclear Multi-Messenger Astrophysics (NMMA), to perform the kilonova analysis. NMMA infers the parameters of the astrophysical transient's source $\theta$ via Bayesian statistics. Based on Bayes' theorem, the posterior $p(\theta | d, \mathcal{H})$ on parameters $\theta$ under a generative hypothesis $\mathcal{H}$ conditioned on data $d$ is given by
\begin{equation}
    p(\theta | d, \mathcal{H}) = \frac{p(d | \theta, \mathcal{H}) p( \theta | \mathcal{H} ) }{p(d | \mathcal{H} )},
\end{equation}
where $p(d | \theta, \mathcal{H})$, $p(\theta | \mathcal{H})$, and $p(d | \mathcal{H})$ are the likelihood, prior, and the evidence, respectively.

The evidence acts as a normalization constant for the posterior, marginalizes the likelihood over the prior,
\begin{equation}
    p(d | \mathcal{H}) = \int d\theta p(d | \theta, \mathcal{H}) p(\theta | \mathcal{H}).
\end{equation}

To compare two hypotheses, $H_{\textrm{A}}$ and $H_{\textrm{B}}$, odds ratio $\mathcal{O}^{\textrm{A}}_{\textrm{B}}$ between them can be used, which is defined as
\begin{equation}
    \mathcal{O}^{\textrm{A}}_{\textrm{B}} = \frac{p(d | H_{\textrm{A}})}{p(d | H_{\textrm{B}})}\frac{p(H_{\textrm{A}})}{p(H_{\textrm{B}})} \equiv \mathcal{B}^{\textrm{A}}_{\textrm{B}}\Pi^{\textrm{A}}_{\textrm{B}},
\end{equation}
where $\mathcal{B}^{\textrm{A}}_{\textrm{B}}$ and $\Pi^{\textrm{A}}_{\textrm{B}}$ are the Bayes factor and the prior odds. If $\mathcal{O}^{\textrm{A}}_{\textrm{B}} > 1$, $H_{\textrm{A}}$ is more plausible than $H_{\textrm{B}}$, vice versa. Throughout this study, the prior odds are set to 1. Therefore, the Bayes factor is used interchangeably with the odds ratio.

The sampling of the posterior distribution and the estimation of the evidence are done by using the nested sampling algorithm~\citep{skilling2006} implemented in \texttt{pymultinest}~\citep{Buchner:2014nha,Feroz:2008xx}.

In this study, we will be comparing the evidence for four different kilonova models developed by Los Alamos National Labs (LANL) using SuperNu~\citep{wollaeger13,wollaeger14} and another model created using the POSSIS radiative transport code~\cite{bulla19,Bulla2023}.
ML architectures within NMMA are trained to produce surrogate models (also called emulators) of light curves that map binary and ejecta parameters to time-dependent magnitudes as quickly as possible, for an arbitrary set of input parameters within a given range.
While advanced radiative transport simulations require days or sometimes months to complete, the use of ML-based emulators not only allows to interpolate a coarse grid of simulations to the entire parametric domain, but also drastically reduces computational resources needed.
Moreover, because NMMA needs to evaluate a light curve millions of times in the course of Bayesian analysis, an emulator is absolutely necessary for speeding up the entire inference process.

NMMA features a variety of kilonova model emulators, ranging from simple diffusion-based semianalytic models \citep{metzger10} to sophisticated radiative transfer ones such as models computed with the POSSIS code \citep{bulla19}, and now also the newly added SuperNu models~\citep{wollaeger21}.
Thereupon, given a set of parameters drawn from a prior distribution, the emulator will generate a light curve template which is matched against data and analyzed to evaluate a likelihood value.
Each ML surrogate model studied below is constructed of several independently trained emulators for each filter included in the host of telescopes used for transient observations.

\begin{table}[h]
\centering
\begin{tabular}{ c | c }
    $\log_{10} M_{\rm ej}^\text{dyn}$ ($M_{\odot}$) & -3, -2.5, -2, -1.5, -1 \\
    \hline
    $v_{\rm ej}^\text{dyn}$ (c) &  0.05, 0.15, 0.2, 0.3 \\ 
    \hline
    $\log_{10} M_{\rm ej}^\text{wind}$ ($M_{\odot}$) & -3, -2.5, -2, -1.5, -1 \\
    \hline
    $v_{\rm ej}^\text{wind}$ (c) &  0.05, 0.15, 0.2, 0.3 \\
\end{tabular}
\caption{ Table of parameter values used for the creation of the grid of light curves from SuperNu.\\}
\label{tab:params}
\end{table}

We use a grid of kilonova light curves computed with SuperNu resulting from a few potential axisymmetric geometries of ejected material~\citep{korobkin21}.
SuperNu is a Monte Carlo radiative transport code which employs an advanced suite of atomic opacities \citep{fontes20}, reflecting detailed elemental composition of the ejecta. 
The ejecta in these simulations is formed of two overlayed components, a lanthanide-rich toroidal component (dynamical ejecta), and a lanthanide-free (wind) component either around the poles, or spherically symmetric.
While we refer to the lanthanide-rich component as the ``dynamical ejecta'' and the lanthanide-free component as the ``wind'',  these terms are used only as labels; it is known that ejecta expelled dynamically could contain lanthanide-free material, for example from the merger interface \citep[the $i$-ejecta,][, their Fig.2]{korobkin12}, while accretion disk wind can be very neutron-rich and produce lanthanide-rich material~\cite{miller19b}.

SuperNu model parameters are summarized in Table~\ref{tab:params}: mass and average velocity of the dynamical ejecta ($M^{\rm dyn}_{\rm ej}$ and $v^{\rm dyn}_{\rm ej}$), mass and average velocity of the wind ($M^{\rm wind}_{\rm ej}$ and $v^{\rm wind}_{\rm ej}$).
Additional parameters are the starting electron fraction of neutron-rich matter as a proxy for nucleosynthesis and composition ($Y_e$), and the inclination angle ($\iota$) between the binary orbital plane and the line of sight.
Two wind compositions are studied, labeled `wind2' and `wind1', with the starting electron fractions $Y_e = 0.27, 0.37$ referred in the TP2/TS2 and TP1/TS1 morphologies respectively.
Composition for the dynamical ejecta and the two types of wind is the same as in Table~2 of ~\cite{wollaeger21}.
The masses ($M_{\rm ej}^\text{dyn}$ and $M_{\rm ej}^\text{wind}$) span $0.001 - 0.1\ M_{\odot}$, and the expansion velocities $v_{\rm ej}^{\rm dyn}$, $v_{\rm ej}^{\rm wind}$ span $0.05-0.3\ c$ (we supplemented the original grid of ~\cite{wollaeger21} with an additional value of $0.2\ c$). 
With spectra from each of 54 different angular bins and each ejecta parameter explained above, this results in a training set of 12150 light curves.

\begin{figure}
    \centering
    \begin{tabular}{cc}
    \includegraphics[width=0.42\columnwidth]{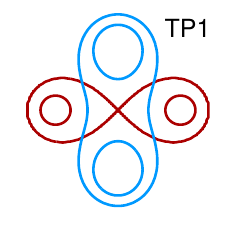} &
    \includegraphics[width=0.42\columnwidth]{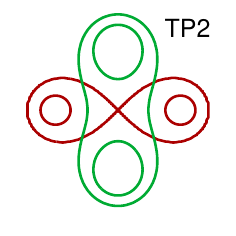}
    \\
    \includegraphics[width=0.42\columnwidth]{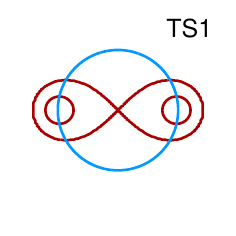} &
    \includegraphics[width=0.42\columnwidth]{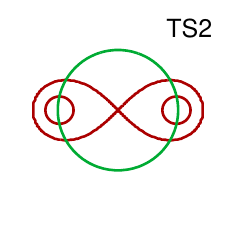}
    \end{tabular}
    \caption{A depiction of the four axisymmetric morphologies used in this study, representing contours of the density cut in the $xz$-plane. 
    In the model notation, the first letter (`T') stands for toroidal dynamical ejecta, the second letter (`S' or `P') for spherical or peanut-shaped wind, and the last digit (2 or 1) stands for the type of wind, with the initial electron fractions $Y_e = 0.27$ and $0.37$, respectively.}
    \label{fig:morphs}
\end{figure}

These models were originally created to test how the kilonova depends on the observed system inclination angle, motivated by simulations of neutron star mergers and postmerger accretion disk which showed highly non-spherical morphologies of the ejecta \citep{rosswog13,miller19b}.
A handful of representative axisymmetric morphologies were created from families of Cassini ovals \citep[see][]{korobkin21}.
Each simulation uses two-component uniform-composition distributions with of low-$Y_e$ (``dynamical ejecta'') and high-$Y_e$ (``wind'') material.
The models are labeled by two letters and a number; the first letter (`T') denotes morphology of the dynamical ejecta component (T = toroidal in all cases), the second labels the morphology of the wind (`S' or `P'), and the number (1 or 2) labels the type of wind.
`S' represents a spherical geometry and `P' represents a peanut shape.
The four classes of models are pictured in the Figure~\ref{fig:morphs}.

These simulations make a valuable addition to the NMMA toolkit, for several reasons.
For once, they are constructed using the realistic suite of atomic opacities, incorporating atomic calculations for several hundred thousands of levels and four ionization stages for the lanthanides~\cite{fontes15a,fontes20}.
Better data for the lanthanides increases the accuracy of our light curves in the optical and IR bands.
Secondly, these models adopt more complex two-component morphologies that increase the dependence of viewing angle.
This allows for greater ability to discern ejecta parameters when the inclination angle of the binary is known from additional evidence.
Finally, these models implement detailed heating energy from $r$-process nucleosynthesis, as well as its time- and density-dependent thermalization, following energy partitioning of different radioactive species (electrons, alphas, and gammas)~\citep{wollaeger18}.

For comparison, we choose the grid of kilonova models created in 2019 using the POSSIS code~\citep{bulla19}.
The comparison is apt because the POSSIS code also implements detailed Monte Carlo radiative transport for computing KN and supernova spectra, and it also uses a two-component uniform matter distribution with lanthanide-rich and lanthanide-poor regions.
The POSSIS models delineate between the lanthanide-rich and lanthanide-poor regions with a parameter called the half-opening angle at the merger plane.
This angle corresponds to half the angular span of the lanthanide-rich region surrounding the orbital plane of the binary \citep[See Figure~1 in][]{bulla19}.

%%%%%%%%%%%%%%%%%%%%%%%%%%%%%%%%%%%%%%%%%%%%%%%%%%%%%%%%%%%%%%%%%%%%%%%%%%%%%%
\section{Results}
\label{sec:results}
\subsection{Verification of Surrogate Model}
\label{subsec:validation}
With any ML-based approach, it is important to verify the training of the emulators, so we performed three tests to assess fitness:
\begin{itemize}
    \item Parameter recovery by applying the NMMA Bayesian inference pipeline on training data.
    \item Analyzing light curves generated by SuperNu that were off the grid points of the original training set.
    \item Determining whether analysis can differentiate between the different morphological model classes.
\end{itemize}

We begin by verifying the surrogate models that we use later in Section \ref{subsec:at2017gfo} to analyze the photometry of the kilonova AT2017gfo.
These are models trained for all four of the morphologies and in each of the filters used in that detection: the `ps1' \emph{grizy} filters \citep{chambers2019panstarrs1surveys}, and the `2mass' $H$, $J$, and $K_s$ filters \citep{Jarrett_2000}.
In the course of analysis, we found the morphology best matching AT2017gfo to be TP2 (see Section \ref{subsec:at2017gfo}), so validation is only shown for the TP2 morphology.

We validate the parameter recovery of a surrogate model by testing if it can accurately find the parameters used to produce a light curve.
This is done applying the NMMA framework to a light curve picked from the training suite.
Note that this test is not trivial, because a light curve from an emulator deviates from the training light curve with the same input parameters.

\begin{figure}
    \centering
    \includegraphics[width=\columnwidth]{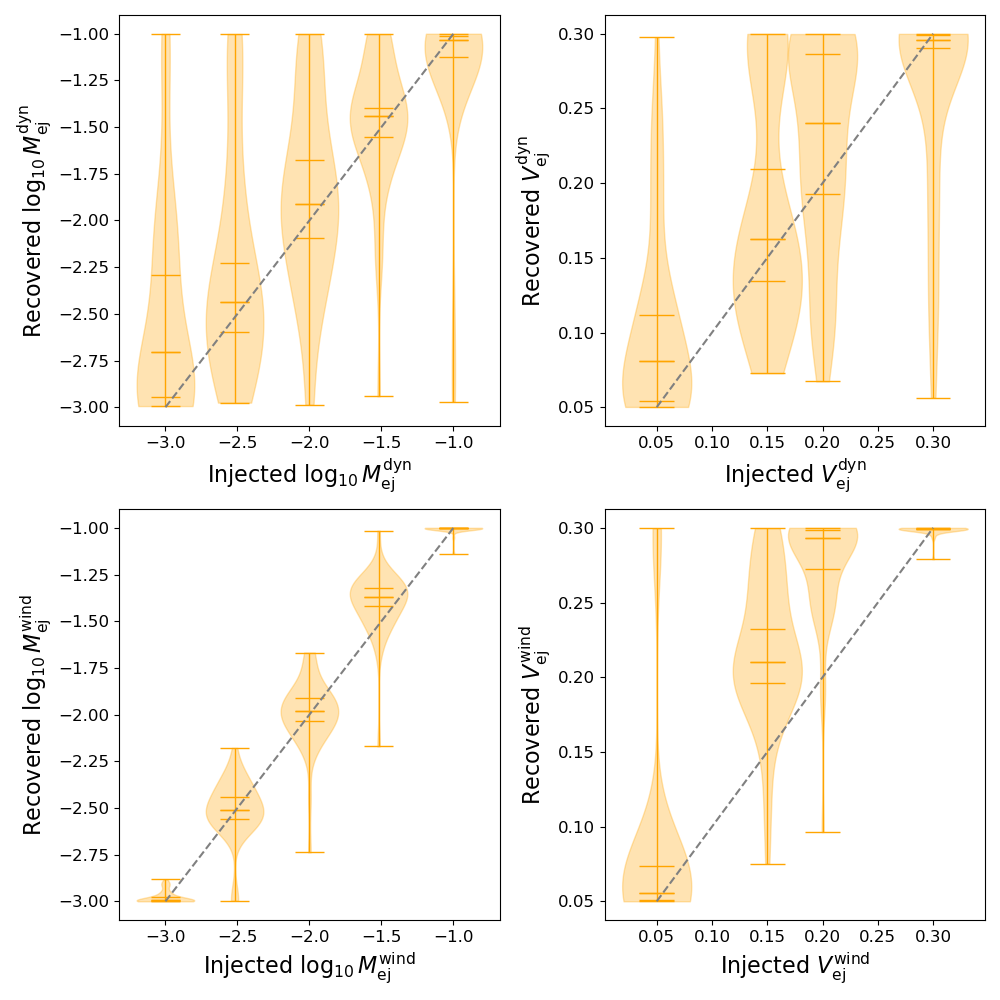}
    \caption{Injected vs.~recovered values for ejecta parameters, clockwise from top left: dynamical ejecta mass ($M_{\rm ej}^{\rm dyn}$), dynamical ejecta velocity ($v_{\rm ej}^{\rm dyn}$), wind velocity ($v_{\rm ej}^{\rm wind}$), and wind mass ($M_{\rm ej}^{\rm wind}$). 
    The violins represent one-dimensional marginal posterior distributions of recovered parameters for the corresponding injection bin. 
    If the surrogate models were perfect in recovering the training data, every recovery would lie along the dashed gray diagonal.}
    \label{fig:TP2NN6}
\end{figure}

Any light curve from the training set is in absolute magnitude so it is adjusted to the apparent magnitude for a kilonova at $D= 40.7\text{ Mpc}$ and nothing else about the data is changed.
We perform Bayesian inference on each test light curve and extract the best fit parameters from the output posterior distributions.
This is repeated for just under 800 random light curves from the 12150 training light curves and the results compiled in the Figure~\ref{fig:TP2NN6}.
With parameter recovery expected to not depend of viewing angle, the number 800 was chosen to have on the order of 150-200 light curves in each parameter bin or approximately $7\%$ of the training set.
True values are along the $x$-axis and the predicted values corresponding to the compiled maximum likelihood estimates are shown as distributions along the $y$-axis.
An ideally trained model would have the recovered parameter distributions centered on the diagonal (shown as a gray dashed line) with zero spread.

While the wind ejecta parameters are recovered with acceptable accuracy, it is clear that this surrogate model struggles to accurately recover the parameters associated with the dynamical ejecta.
The trained model does better recover the wind ejecta parameters, notably the wind mass, however overestimating the wind velocity in the middle of the parameter space.
We speculate this overestimation is a consequence of the surrogate model being unable to accurately capture (relatively) large and irregular oscillations in magnitudes during early times.
Next, we test how proposed JWST follow-up (studied later Section \ref{subsec:morph}) affects validation of a surrogate model trained with additional inclusion of the `f560w' and `f770w' filters planned for the MIRI imaging~\citep{Dicken_2024}.
Apart from the training now covering longer time span, the validation is identical to the process described above but with the inclusion of the two extra points from infra-red bands.

\begin{figure*}
    \centering
    \begin{tabular}{cc}
    \includegraphics[width=0.47\textwidth]{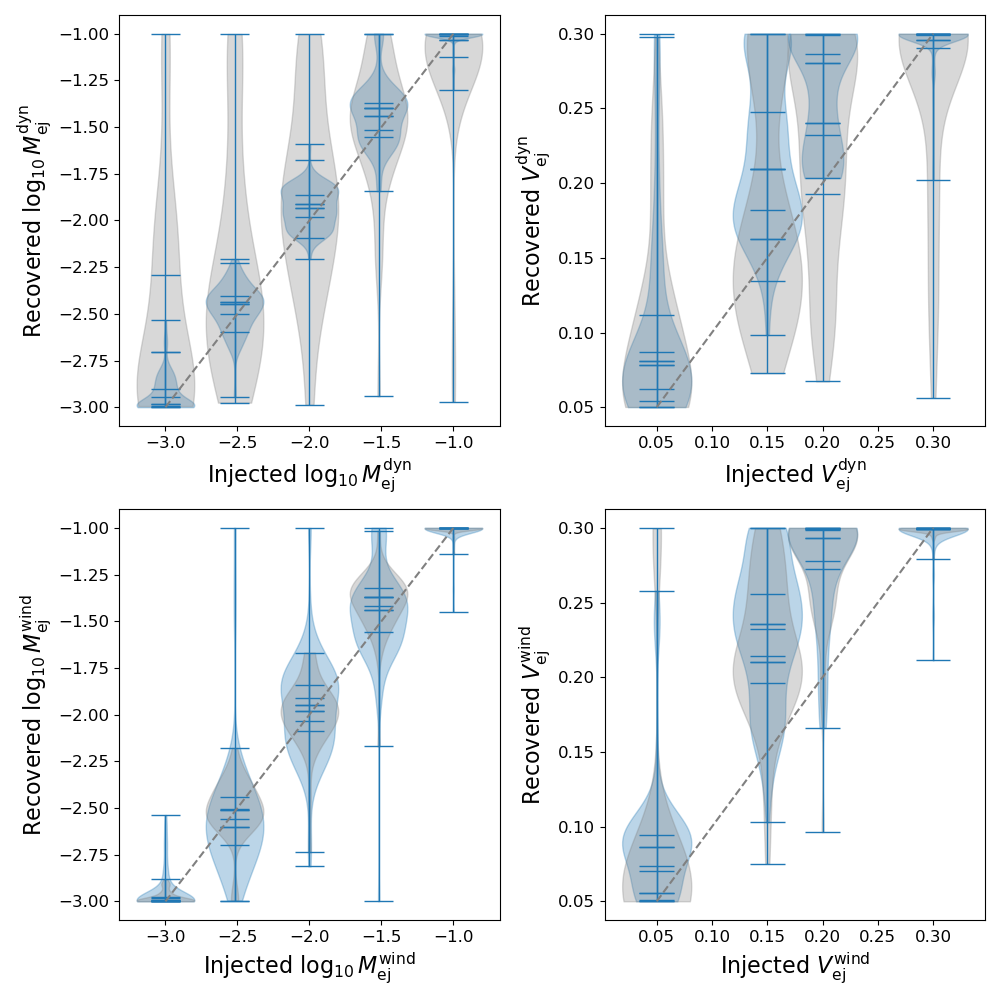} &
    \includegraphics[width=0.47\textwidth]{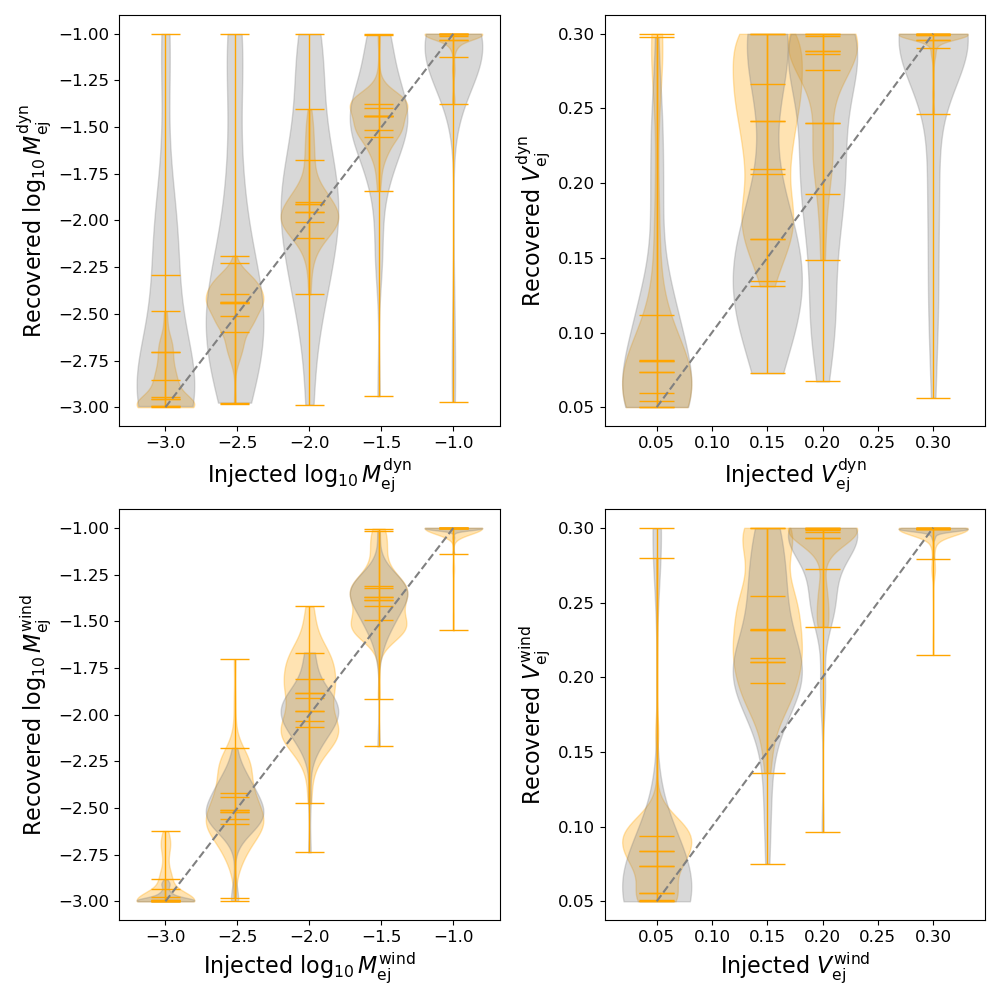}
    \\
    \includegraphics[width=0.47\textwidth]{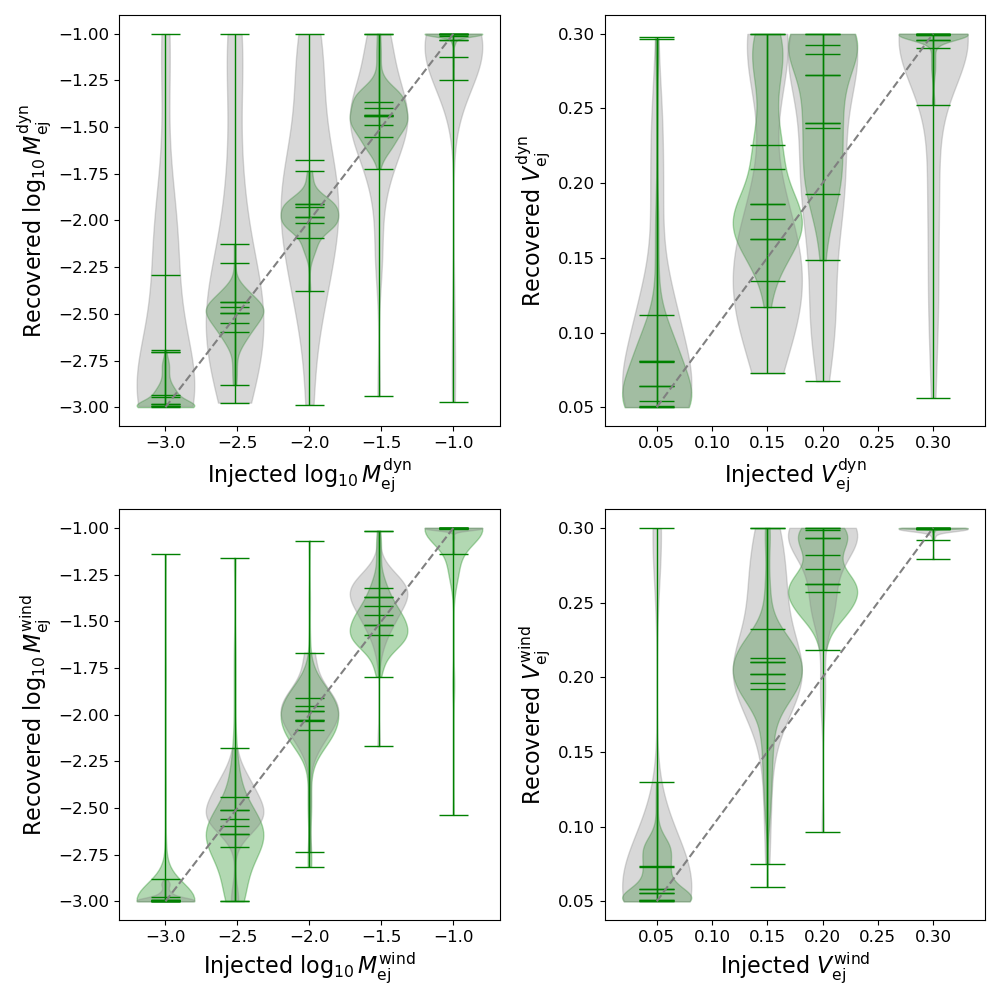} &
    \includegraphics[width=0.47\textwidth]{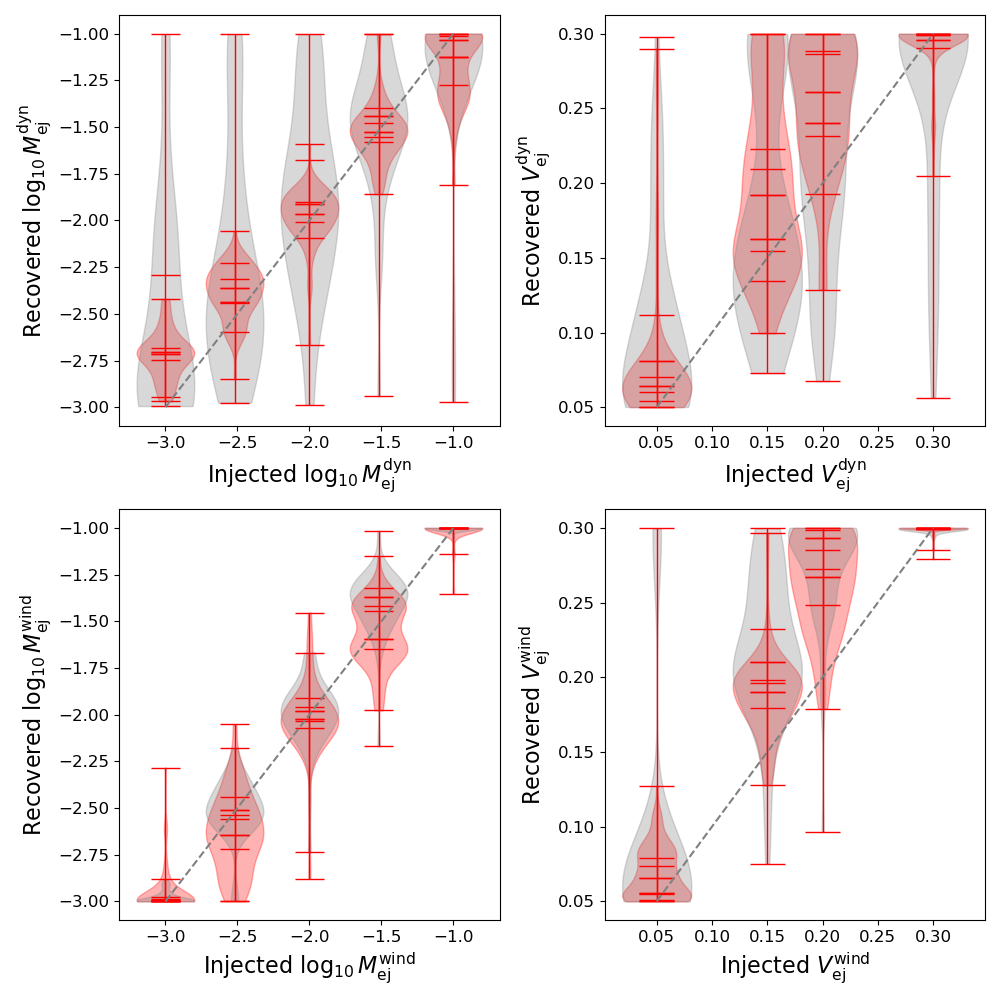}
    \end{tabular}
    \caption{Comparison of injected vs. recovered values for ejecta parameters for each morphology compared to the TP2 morphology (gray), also shown in Figure \ref{fig:TP2NN6}. Violins show marginal one-dimensional posterior distributions for 800 injections. The parameters in each plot are: (clockwise from top left) dynamical ejecta mass, dynamical ejecta velocity, wind velocity, and wind mass. 
    Different morphologies are color coded through the rest of the analysis: TP1 is blue, TP2 is orange, TS1 is green, and TS2 is red.
    The gray dashed diagonal line represents the perfect recovery, if every input was mapped to the correct output parameter. Some deviation from this diagonal is expected with a trained model that is not over-tuned.}
    \label{fig:jwst_val}
\end{figure*}

Shown in Figure~\ref{fig:jwst_val} are surrogate recovered-vs-injected distributions for each morphology \emph{with} additional JWST photometry, overlayed with the TP2 surrogate model validation results \emph{without} JWST photometry (the one in Figure~\ref{fig:TP2NN6}).
Of note is how much better the surrogate models recover the dynamical ejecta properties with the inclusion of MIR filter data, which shows potential for improved kilonova discovery in the JWST era.
This is due to the fact that the dynamical ejecta is much brighter than the wind ejecta in the IR than in optical bands at late times (one-two weeks) due to its high lanthanide fraction.

The next step in validating the surrogate models is testing how well the models can interpolate the light curves and recover ejecta parameters from the data produced by off-grid parameters (but within the ranges of the parameter grid).
To quantify this, we calculate the average chi-squared deviation between the SuperNu light curve and the surrogate-generated light curve.
Then, we use the off-grid light curves as data input for the analysis pipeline and compare true parameters to recovered parameters.

We use a total of twelve off-grid validation light curves produced with three sets of input parameters for each of the four morphological models.
The summary of the input parameters, chi-squared, and recovered parameters are found in Table \ref{tab:offgrid}.
Most of the light curves result in a reasonable reduced chi-squared ($\chi^2_\nu \simeq 1$) assuming an error in the surrogate models of $\sigma=0.4$ in units of magnitude.
Also interesting to note is the reduction of all chi-squared values with the inclusion of the IR filter data.
Because the surrogate model for each filter is trained independently, the deviation of the emulator from the off-grid light curves is lower for the IR bands than all other bands.

\begin{table*}[htb]
\centering
\begin{tabular}{ c c | c  c | c  c | c  c | c  c | c  c}
    (\#) & Morphology &  $M_\text{in}^\text{dyn}$ &  $M_\text{out}^\text{dyn}$ ($M_{\odot}$)  & $M_\text{in}^\text{wind}$ & $M_\text{out}^\text{wind}$ ($M_{\odot}$) & $v_\text{in}^\text{dyn}$ & $v_\text{out}^\text{dyn}$ (c) & $v_\text{in}^\text{wind}$ & $v_\text{out}^\text{wind}$ (c) & $\chi^2_\nu$ & $\chi^2_\text{jwst}$\\
    \hline
(i)   & TP1 & 0.085 & 0.100 & 0.013 & 0.019 & 0.231 & 0.262 & 0.251 & 0.294 & 1.6 & 1.4 \\
(ii)  & TP1 & 0.001 & 0.002 & 0.008 & 0.009 & 0.099 & 0.265 & 0.247 & 0.289 & 2.3 & 1.0 \\
(iii) & TP1 & 0.070 & 0.097 & 0.007 & 0.003 & 0.116 & 0.167 & 0.260 & 0.292 & 4.5 & 3.2 \\
(iv)  & TP2 & 0.002 & 0.003 & 0.081 & 0.100 & 0.180 & 0.282 & 0.131 & 0.167 & 8.8 & 3.9 \\
(v)   & TP2 & 0.011 & 0.017 & 0.005 & 0.005 & 0.259 & 0.300 & 0.123 & 0.176 & 4.2 & 2.7 \\
(vi)  & TP2 & 0.005 & 0.007 & 0.003 & 0.002 & 0.263 & 0.299 & 0.281 & 0.298 & 6.3 & 2.3 \\
(vii) & TS1 & 0.064 & 0.100 & 0.008 & 0.015 & 0.058 & 0.075 & 0.164 & 0.096 & 12.6 & 8.6 \\
(viii)& TS1 & 0.016 & 0.024 & 0.003 & 0.003 & 0.186 & 0.276 & 0.201 & 0.156 & 9.1 & 2.8 \\
(ix)  & TS1 & 0.026 & 0.025 & 0.011 & 0.019 & 0.286 & 0.300 & 0.077 & 0.063 & 7.0 & 1.6 \\
(x)   & TS2 & 0.002 & 0.003 & 0.039 & 0.035 & 0.154 & 0.159 & 0.192 & 0.117 & 2.3 & 2.7 \\
(xi)  & TS2 & 0.014 & 0.020 & 0.007 & 0.005 & 0.290 & 0.297 & 0.092 & 0.067 & 12.6 & 8.6 \\
(xii) & TS2 & 0.004 & 0.005 & 0.017 & 0.015 & 0.090 & 0.101 & 0.222 & 0.146 & 8.8 & 5.3 
\end{tabular}
\caption{This table shows the average reduced chi-squared values for each off-grid light curve, their associated ejecta parameters, and the parameters recovered from the analysis pipeline. Three chi-squared values are shown, the first trained model, that model without the `ps1' g-band, and the model including JWST filters. The chi-squared deviations are averaged over all filters and 54 observing angles. The recovered parameters are only averaged over all 54 observing angles. \\}
\label{tab:offgrid}
\end{table*}

\begin{figure}
    \centering
    \includegraphics[width=\columnwidth]{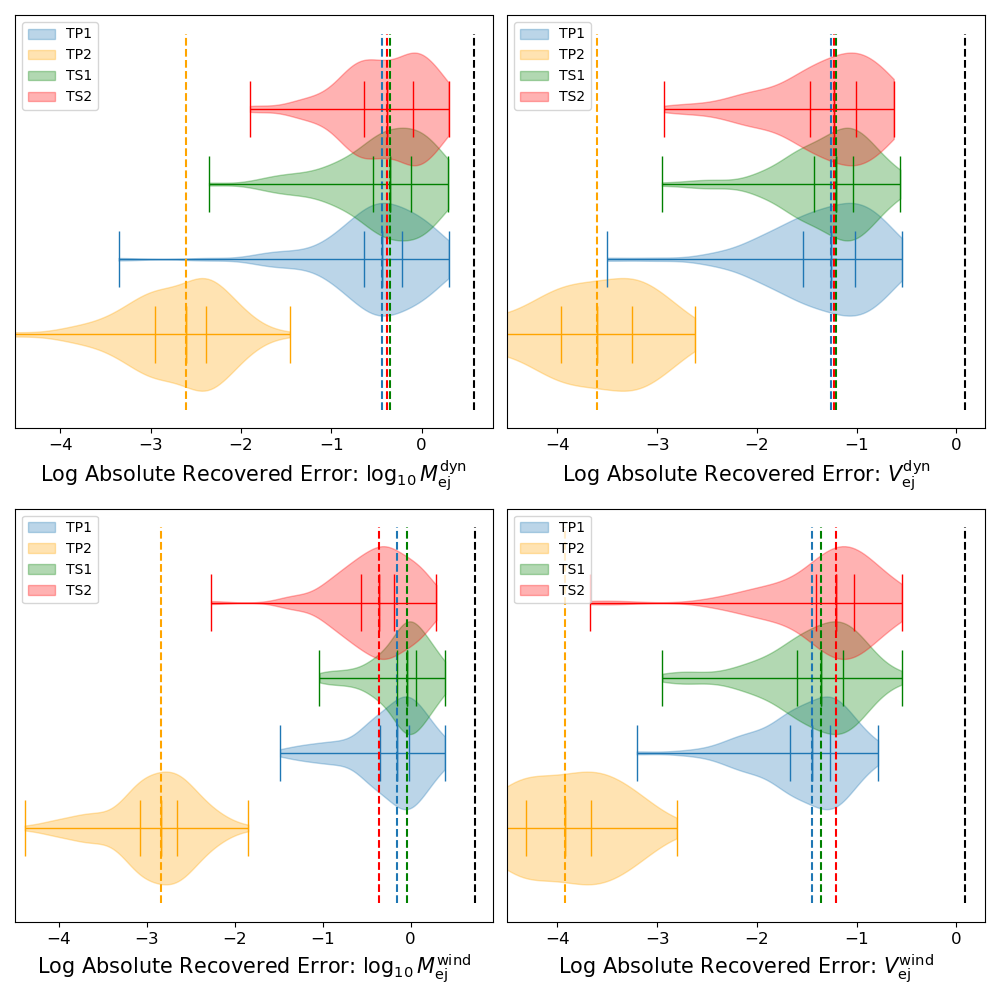}
    \caption{Violin plots showing the distribution of absolute errors between injected and recovered parameters for 100 light curves injections, where injections are performed with all four morphologies, and recovery is done assuming TP2. 
    The error is summed over 56 time steps in each of 27 filters going out to 14 days.
    The four panels are for the ejecta parameters (in clockwise order): dynamical ejecta mass, dynamical ejecta velocity, wind velocity, and wind mass. 
    The colored dashed lines indicate the median errors for the different morphologies and the solid vertical lines are the bounds and 1$\sigma$ (67\%) quantiles.
    The black dashed line is the error expected for random sampling of the inferred parameter from the prior distribution. Because all recoveries are done using the TP2 morphology, the absolute error of input light curves constructed using the TP2 emulator should be the lowest if light curves from each morphology is distinct. }
    \label{fig:abs_err}
\end{figure}

The final step in validation was checking if each trained morphological model was distinct, i.e. they only correctly recover parameters from light curves generated by that model.
To do this, we generated 100 light curves from each morphology (TP1, TP2, TS1, TS2) and ran inference on those light curves assuming a TP2 model.
If there are distinct classes of light curves associated with each morphology, the light curve parameters should only be correctly recovered using the TP2 model, and we see exactly this in Figure~\ref{fig:abs_err}.
This Figure shows absolute error distributions in recovery of the ejecta parameters.
We calculate the error between the injected and recovered ejecta parameters with the 90\% confidence intervals of the posteriors as error bars for each of the four trained models.
We measure the deviation of the recovered parameters for each injection from the injected values and show the median error by the dashed line.

It follows from the results presented in Figure~\ref{fig:abs_err} that if the light curve has the characteristics of a TP2 morphology, inference with TP2 correctly predicts the ejecta parameters.
The TP2 model (orange) has its median error 1-2 orders lower than the median error the other morphologies.
We point out that all models perform about an order of magnitude better than a random guess in the prior space, shown by the black dashed line.
This illustrates a simple fact that models with wrong morphology can still recover ejecta parameters, just mot as accurately as the model with the right morphology.
Because the TP2 model is best recovered by the analysis using TP2 emulator, it shows that the emulators for each morphology are distinct classes inside our ML implementation.
This is important because it means, if the data is sufficient, there is no misinterpretation of one morphology as another.

There are two primary sets of surrogate models used in this project: The TP2 model trained on the filters used in the detection of AT2017gfo and the set of four morphologies trained with the inclusion of two filters used in the planned follow-up by JWST.
Each morphological model is distinct as shown by Figure~\ref{fig:abs_err} meaning, on average, any reliable recovery of one morphology is not a misinterpretation of a light curve from another model given there is enough data.
All of the surrogate models are accurately interpolating light curves in between the grid of ejecta parameters used for training.
This is supported by the reduced chi-squared deviations listed in Table \ref{tab:offgrid} and the recovery of parameters for the off-grid light curves that are within the on-grid recovery distributions shown in Figure \ref{fig:jwst_val}.
Recovery of input parameters for the TP2 surrogate used for analysis of AT2017gfo is sufficient for the wind mass parameter but there is little to no discerning power for the dynamical ejecta parameters.
Furthermore, this surrogate model shows a systematic over-estimation of the wind velocity parameter.
However, the model trained with the inclusion of the IR band data from JWST does much better in these regards.

\subsection{Realistic Determination of Morphology}
\label{subsec:morph}
Testing the disjointedness of the morphological classes above (Section \ref{subsec:validation}) we injected light curves with fifty data points in every filter, however this is an extremely optimistic case for a kilonova observation.
The serendipitous observation of AT2017gfo (the kilonova that accompanied GW170817) used in the analysis of Section \ref{subsec:at2017gfo} only had $\sim$~30 data points total over eight filters.
This is near the best possible observation of a kilonova we can get at this time and other observations are expected to be even more data-poor.
With this in mind, we tested if more realistic kilonova observing cadences could adequately determine the morphology of the ejecta.
We looked at the cadences for LSST in the new Vera Rubin Observatory ~\citep{rubin24} which is expected to come online in early 2026 along with proposed observing scenarios with the James Webb Space Telescope (JWST).

The Rubin observatory uses the LSST instrument which observes transients in up to five bands and at deeper limiting magnitudes than previous kilonova surveys (such as ZTF).
Similarly to the testing of disjoint models above, we analyzed 100 simulated light curves from our emulators as data and recovered ejecta parameter values for each model using the NMMA package.
Instead of 50 data points in 23 different filters, we used a cadence specified by the observing strategy documentation ~\citep{rubin24}.
Target of Opportunity (ToO) observations of kilonovae with the Rubin Observatory will have three modes but we focused on the most optimistic case, when gravitational wave detections have localized the event to under 30 square degrees.
If this requirement is met, Rubin will observe that patch of sky a total of six times in the `ps1' $grizy$ filters; Three times on the first night the trigger is sent and one time each for three subsequent nights.

\begin{figure}
    \centering
    \includegraphics[width=\columnwidth]{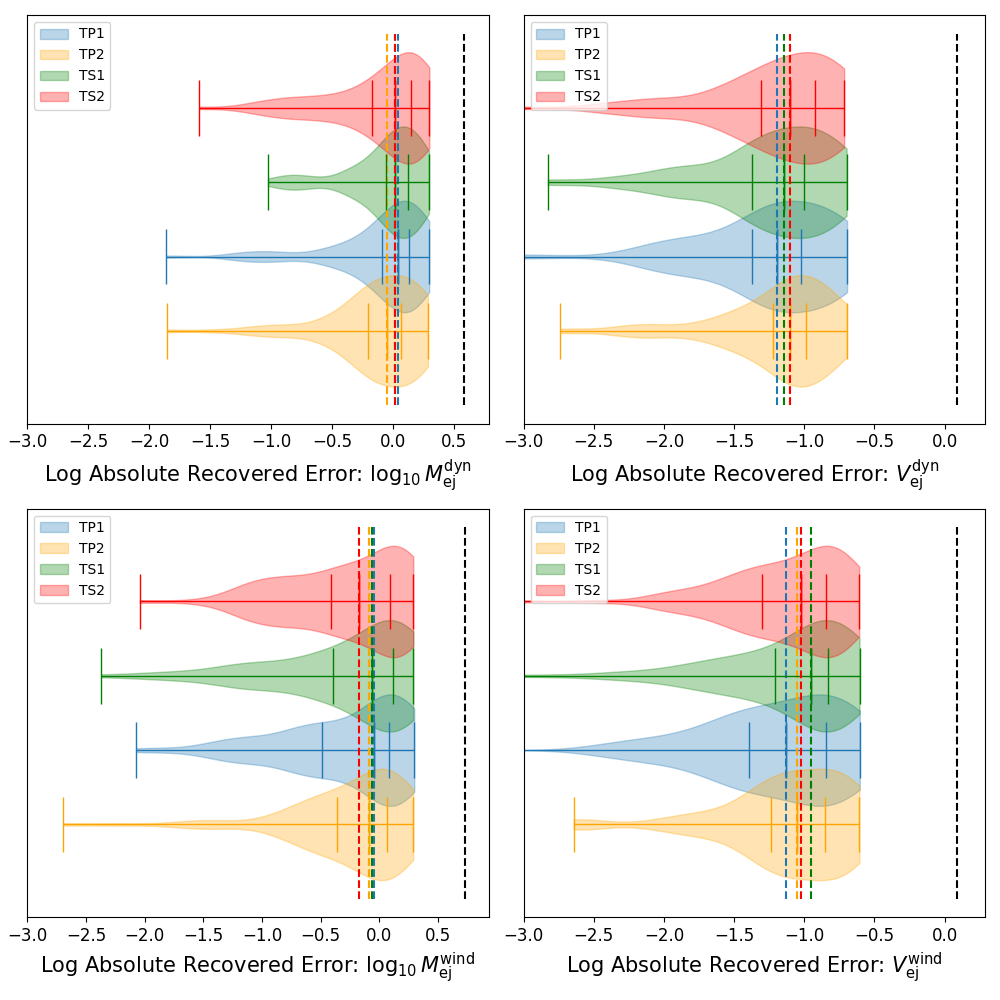}
    \caption{Violin plots showing distributions of absolute errors in parameter recovery for 100 simulated light curves, summed only over the filters and observation points for the planned Rubin follow-up strategy (six observations over four nights) for events well localized by GW observations. 
    Notation is the same as in Figure~\ref{fig:abs_err}.
    No distinction can be made between the morphologies with only the Rubin data.}
    \label{fig:rubin_abs_err}
\end{figure}

A summary of parameter recovery is shown in Figure~\ref{fig:rubin_abs_err} and it is clear that the Rubin ToO plan will not succeed at discerning morphology, as there is no preference for the TP2 morphology in any of the recovered parameters.
However, all models systematically recover masses and velocities of both components better than random guess by almost an order of magnitude in log of the absolute value of error.
The Figure shows that the masses are recovered with an average error of about 1~dex, while velocities with an error of about 0.1~dex.

\begin{figure}
    \centering
    \includegraphics[width=\columnwidth]{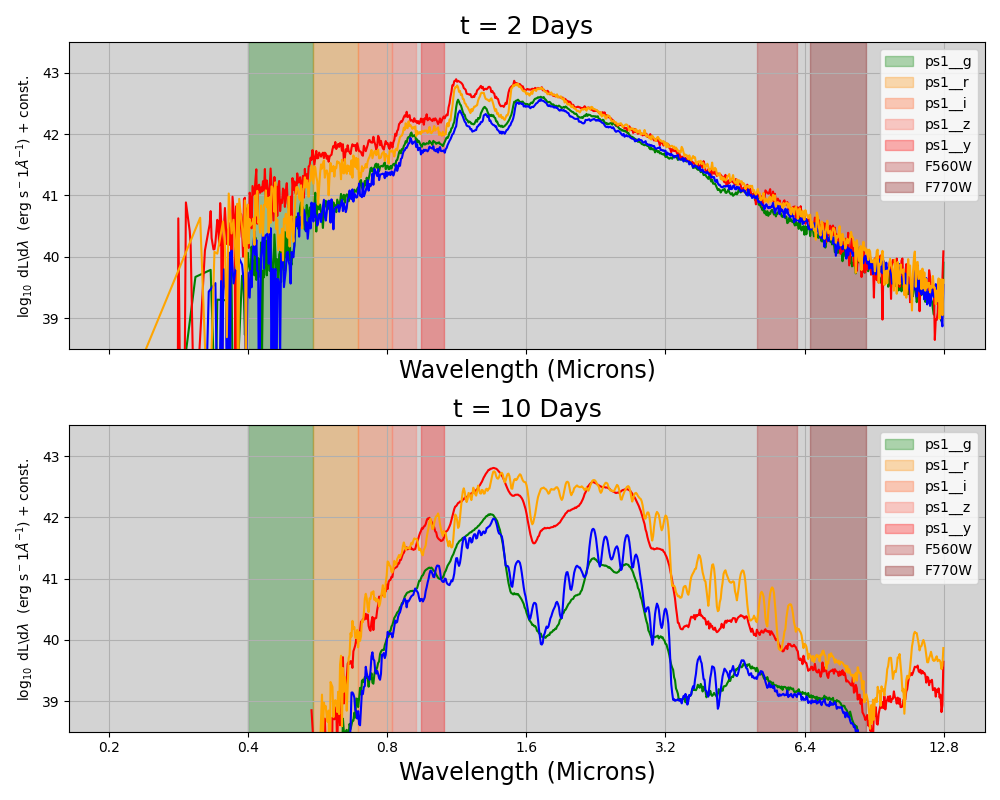}
    \caption{Plotted are sample spectra for the different morphological models (TP1 - blue, TP2 - orange, TS1 - green, TS2 - red) at 2 days (above) and 10 days (below) after merger, with parameters $M_\text{ej}^\text{dyn} = 0.001M_\odot$, $v_\text{ej}^\text{dyn} = 0.3c$, $M_\text{ej}^\text{wind} = 0.1M_\odot$, $v_\text{ej}^\text{wind} = 0.05c$. Overlayed are the bands for the `ps1' filters and the two JWST filters used in the morphology analysis (Section \ref{subsec:morph}). The spectra start as very similar and only deviate substantially. 
    One can also see the similarities between the TS2 and TP2 spectra for this example which explains why the analyses for those two models tend to align. Spectrum plots were produced with the Python scripts based on the Cocteau astro toolkit, \url{https://github.com/eachase/cocteau}.}
    \label{fig:early_vs_late}
\end{figure}

It is curious to compare this result with the recent studies of the P Cygni features in early spectra (up to five days) of AT2017gfo, that strongly suggest the early ejecta had spherical shape \citep{Sneppen_2023}. 
Because the proposed Rubin follow-up strategy only lasts for a maximum of four days, it is improbable we will be able to classify the morphologies of the events from photometry alone.
It is clear that morphological differences are not apparent in the light curve at such early time, however it might be possible to determine morphology from data taken for up to ten days post merger (See Figure~\ref{fig:early_vs_late}).

Other aspects of a multi-messenger signal may help with the prediction of ejecta parameters, so in this next analysis we construct a better scenario for determining morphology when observing a kilonova with LSST.
While the ejecta is close to spherical in the early times post merger, deviations from sphericity can be seen at later times in our models, so instead of simulating observations with the proposed Rubin strategy, we look at the kilonova signal for ten days following the merger.
Inference of parameters can also be helped if we can restrict the inclination angle via an independent observation, so we assume an event with coincident gravitational waves and possible gamma ray burst (GRB).
Considering a three-detector GW event, the sky map localization will likely be under 100 square-degrees due to the triangulation of the waves passing through the detector.
If this is a well localized event, either by gravitational waves and/or gamma ray telescopes it is feasible to observe the probable region of the sky to quickly identify the transient.
This will allow for more time to observe the kilonova signal and extract as much information as possible.
Moreover, if the distance is known by placing the event in a known galaxy and the inclination angle is constrained by the gamma/X-ray off-axis afterglow observations, observing kilonova for up to ten days post merger becomes much more informative.

\begin{figure}
    \centering
    \includegraphics[width=\columnwidth]{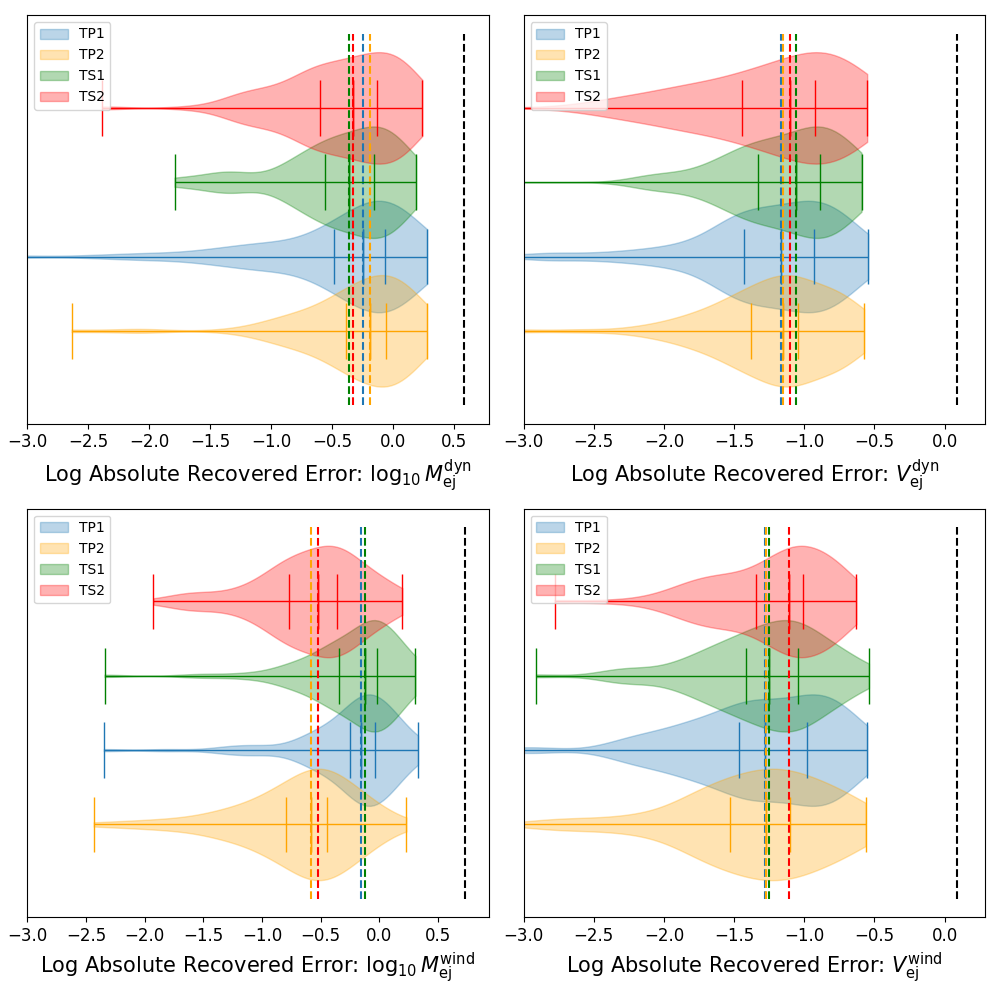}
    \caption{Violin plots showing the distribution of absolute errors for 100 simulated light curves made using an idealized 10 day Rubin follow-up strategy with the distance known from either GW signals or host galaxy attribution. 
    Notation is the same as in Figure~\ref{fig:abs_err}.
    Here the morphology cannot be distinguished, but the type of wind (TP1/TS1 vs TP2/TS2) is separable by the wind mass parameter. 
    The average Bayes factor for these runs heavily favors the TP2 model to TS1 and TP1 ($\mathcal{B}=33$, $48$) and only slightly favors TP2 to TS2 ($\mathcal{B}=1.4$).}
    \label{fig:rubin_10d_fixDl_abs_err}
\end{figure}

With continued observation to ten days and a best case multiple signal event, it is possible to discern morphological differences with photometry only:
as we see in Figure~\ref{fig:rubin_10d_fixDl_abs_err}, the error in parameter recovery for the TP2 model for this strategy is again on average lower than any other model.
Comparing Bayes factors gives the same conclusion, the TP2 model is heavily favored over the TS1 and TP1 models ($\mathcal{B}=33$, $48$) and slightly favored to the TS2 model ($\mathcal{B}=1.4$).
This result leads to a discussion about the proposed Rubin strategy and suggests if we want to better probe the dynamics of the ejecta proceeding merger events, the Rubin strategy should continue to observe the EM counterpart longer than five days.
With the increasing limiting magnitudes of the instrument, it is quite feasible to take data from nearby events for a few days past the proposed four days.
Light curve data at late times will be a boon to groups that do simulations and try to study the effects of non-LTE (local thermal equilibrium) in the ejecta.

Another option to improve the dataset used for analysis is to include data from the new JWST missions.
Because kilonovae are characteristically red they are perfect follow-up targets for the new JWST mission which specializes in observing the universe in the reddest optical and infrared (IR) bands.
We model JWST observations in the study after two proposals recently accepted by the collaboration with three planned observations by the MIRI instrument in the `f560w' and `f770w' filters.
In the two MIRI filters we simulate observations at 8,15, and 20 days post merger.

\begin{figure}
    \centering
    \includegraphics[width=\columnwidth]{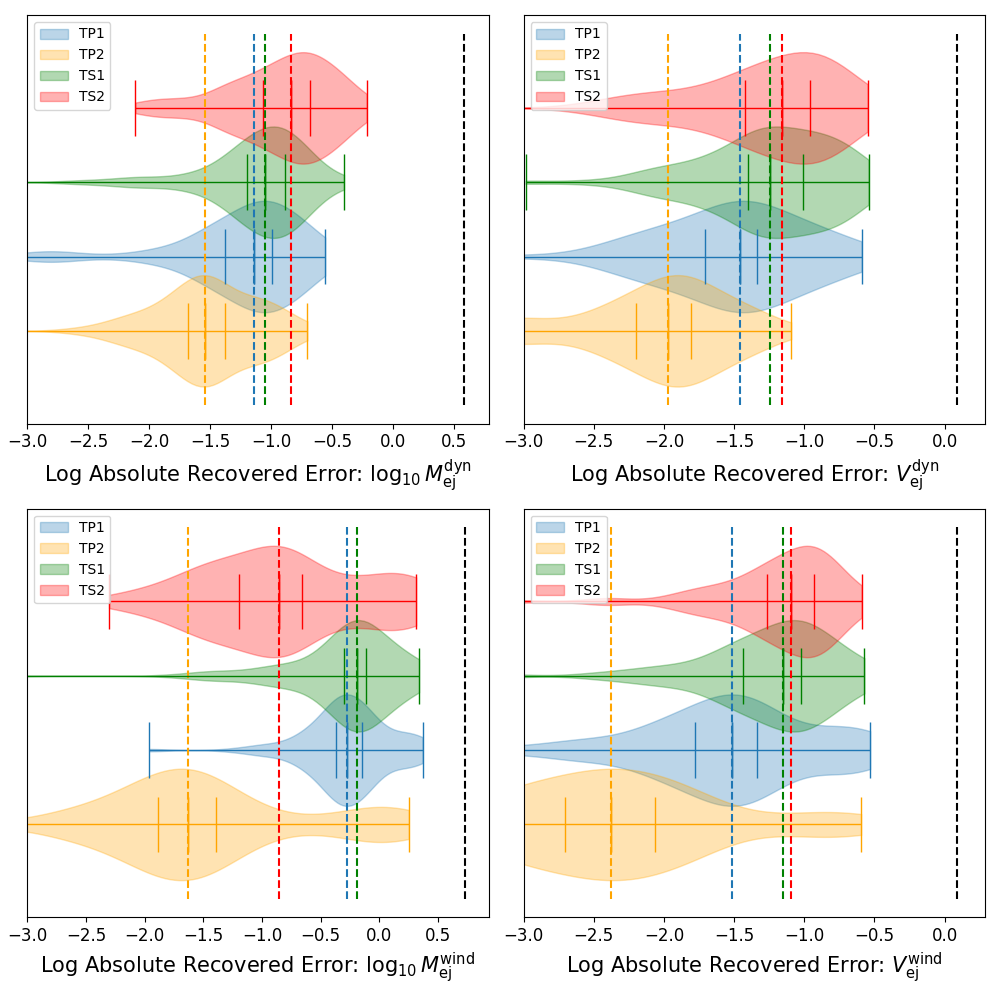}
    \caption{Violin plots showing the distribution of absolute errors for 100 simulations of the four-night Rubin follow-up from Figure~\ref{fig:rubin_abs_err}, supplemented by  three JWST observations on nights 8, 15, and 20. Notation is the same as in Figure~\ref{fig:abs_err}.
    This also assumes a coincident GRB detection, restricting the inclination angle to upper quadrant of the sphere.
    Bayes factors for this case all favor the TP2 morphology ($\mathcal{B}=7.4$, $4.1$, $2.4$ respectively) and the dynamical mass and velocity distributions for the TP2 recovery are distinct from the other morphologies.}
    \label{fig:jwst_rubin_wgrb}
\end{figure}

With the original Rubin observation plan plus the additional three MIRI observations, it is becomes possible to distinguish the morphology of the model, as seen in Figure~\ref{fig:jwst_rubin_wgrb}.
The added MIR photometry data allow the models to distinguish the dynamical mass ejecta much more effectively.
The dynamical ejecta velocity is also better determined due to the late-time additional MIR photometry.
Comparing Bayes factors confirms this conclusion: the TP2 model is favored to all three of the TP1, TS1, TS2 models with $\mathcal{B}=7.4$, $4.1$,$2.4$ respectively.
This means if kilonovae generally have the same morphology, it is possible to determine which morphological model best matches the true ejecta distribution.

\subsection{Analysis of AT2017gfo}
\label{subsec:at2017gfo}
\begin{figure*}
    \centering
    \includegraphics[width=\textwidth]{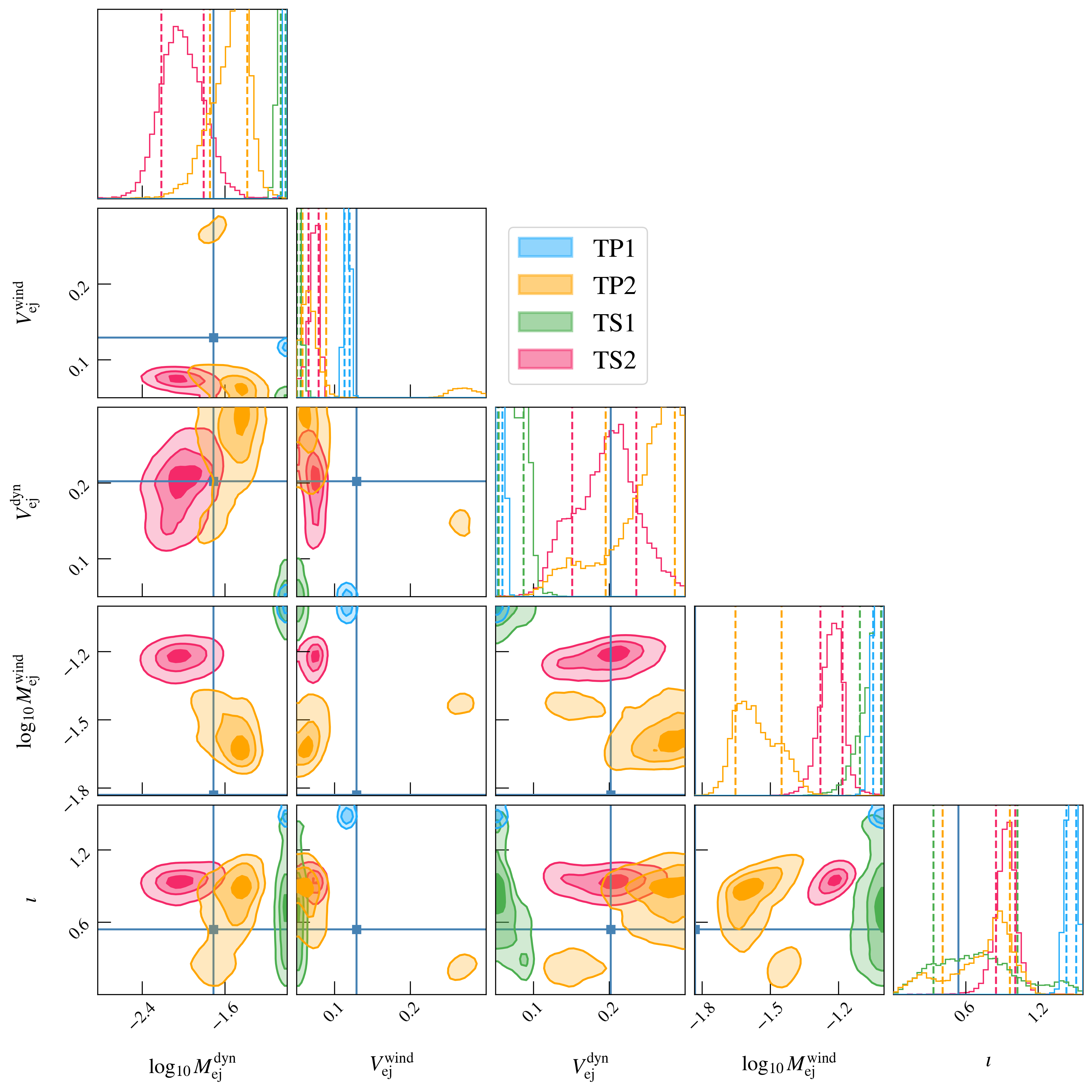}
    \caption{Posterior distributions of recovered kilonova parameters for the four SuperNu-based models resulting from the Bayesian inference of the AT2017gfo photometric observations. 
    First we note that the TP1 (blue) and TS1 (green) models are railing in every ejecta parameter, so those results are discounted because those models are not good fits to the data. Both the TP2 (orange) and TS2 (red) models are reasonable fits to the data. The maximum likelihood values and the 90\% confidence intervals are summarized in Table \ref{tab:newposteriorstats}. The blue points correspond to the values found by \cite{ristic23b}.}
    \label{fig:LANL_comp}
\end{figure*}

\begin{figure*}
    \centering
    \includegraphics[width=\textwidth]{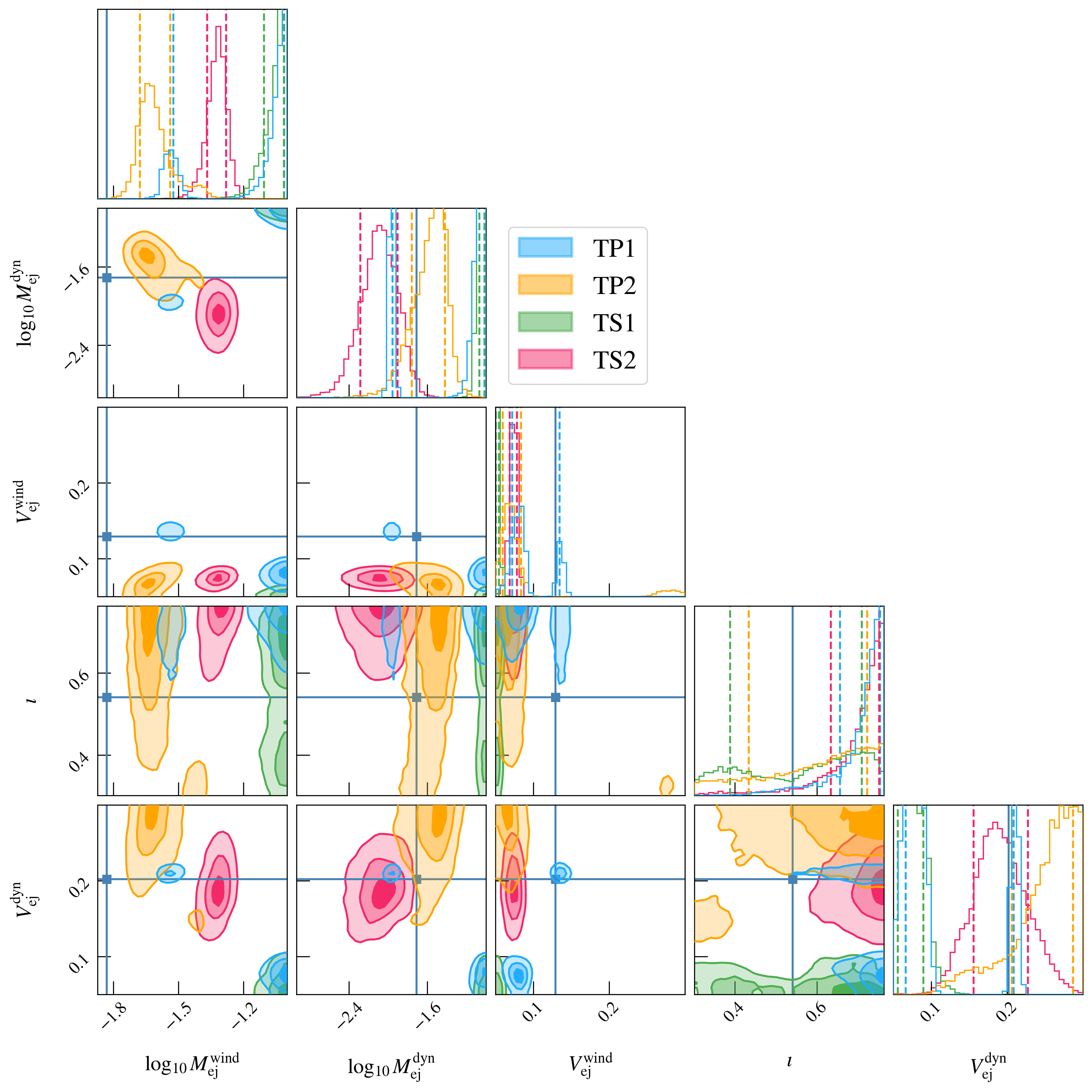}
    \caption{Posterior distributions of recovered kilonova parameters for each of the models resulting from the Bayesian inference of the AT2017gfo light curve where we fix the luminosity distance to $40.7$Mpc and the inclination angle to the range found in \citep{Finstad_2018}. There are again railing issues with the TP1 (blue) and TS1 (green) models, so those models are discounted as good matches to the data. The maximum likelihood values and the 90\% confidence intervals are summarized in Table \ref{tab:newposteriorstats}. The blue points correspond to the values found by \cite{ristic23b}.}
    \label{fig:LANL_comp_iota}
\end{figure*}

We show above that our models form distinct morphological classes and with enough data the ejecta morphology can be recovered.
Here we apply our models to AT2017gfo, the kilonova signal that accompanied GW170817, and determine the morphology that best represents this event.
We performed two analyses using NMMA for each of the four models: one with the free inclination angle parameter $\iota$, and the other with the inclination angle fixed to the range found in ~\citep{hotokezaka2018hubble}.
Both analyses have luminosity distance fixed at $40.7$~Mpc, the best measure for the host galaxy NGC4993 \citep{Hjorth_2017}.
Some future kilonova detections will not have a well known inclination angle, especially if a GRB signal is absent, so we show both cases for posterity.
Posterior distributions from the analyses with the inclination angle free are shown in Figure~\ref{fig:LANL_comp} and results for the run with inclination fixed are presented in Figure~\ref{fig:LANL_comp_iota}.

For the free inclination angle, we find that the TP1 (blue) and TS1 (green) models are ``railing'' (clustering near the boundaries of the parameter range) in all four of the ejecta parameters, and TP1 prefers an inclination angle far from an accepted value from other works.
From this it is clear that the TP1 and TS1 models are poor fits to the kilonova AT2017gfo and it is safe to assume this event did not have ejecta matching these morphologies.
Both the TP2 (orange) and TS2 (red) models give inclination angles near the results of \citep{hotokezaka2018hubble} and \citep{Finstad_2018} and also ejecta parameters that match that of \citep{ristic22}.
However the TP2 model is preferred over the TS2 model because of the combination of wind ejecta parameters.
The high wind velocity combined with the relatively high wind mass means unusually high mass of an accretion disk, which is strongly disfavored by numerical simulations~\cite{fernandez15}.
It is also important to note that the low wind velocity does not fall in the range where the surrogate models systematically over-estimated that parameter, giving us more confidence in the analysis.
The analysis with the inclination angle fixed does not make any noticeable changes to the ejecta parameters, only shifting the peak of the recovered inclination angle within the range in agreement with other works \cite{hotokezaka2018hubble,Finstad_2018}

\begin{table*}[!htb]
\centering
\begin{tabular}{ c | c | c | c | c | c }
    Model &  $M_{\rm ej}^\text{dyn}$ [$M_{\odot}$] & $v_{\rm ej}^\text{dyn}$ [c] & $M_{\rm ej}^\text{wind}$ [$M_{\odot}$] & $v_{\rm ej}^\text{wind}$ [c] & $\iota$ [rad]\\
    \hline
    TP1 & $0.097^{+0.002}_{-0.002}$ ($0.087^{+0.076}_{-0.009}$) & $0.06^{+0.01}_{-0.01}$ ($0.08^{+0.13}_{-0.02}$) & $0.093^{+0.004}_{-0.004}$ ($0.089^{+0.059}_{-0.009}$) & $0.12^{+0.01}_{-0.01}$ ($0.08^{+0.05}_{-0.01}$) & $1.48^{+0.04}_{-0.04}$ ($0.73^{+0.03}_{-0.07}$) \\ [5pt]
    TP2 & $0.031^{+0.011}_{-0.013}$ ($0.028^{+0.010}_{-0.011}$) & $0.26^{+0.03}_{-0.06}$ ($0.26^{+0.03}_{-0.05}$) & $0.027^{+0.005}_{-0.007}$ ($0.024^{+0.005}_{-0.003}$) & $0.07^{+0.02}_{-0.01}$ ($0.06^{+0.01}_{-0.01}$) & $0.81^{+0.16}_{-0.40}$ ($0.62^{+0.10}_{-0.19}$) \\ [5pt]
    TS1 & $0.097^{+0.002}_{-0.007}$ ($0.093^{+0.004}_{-0.008}$) & $0.06^{+0.02}_{-0.01}$ ($0.07^{+0.02}_{-0.01}$) & $0.091^{+0.007}_{-0.012}$ ($0.091^{+0.007}_{-0.012}$) & $0.05^{+0.01}_{-0.01}$ ($0.05^{+0.01}_{-0.01}$) & $0.64^{+0.36}_{-0.34}$ ($0.59^{+0.11}_{-0.21}$) \\ [5pt]
    TS2 & $0.010^{+0.006}_{-0.003}$ ($0.008^{+0.004}_{-0.003}$) & $0.20^{+0.04}_{-0.05}$ ($0.19^{+0.04}_{-0.03}$) & $0.059^{+0.006}_{-0.006}$ ($0.048^{+0.005}_{-0.005}$) & $0.07^{+0.01}_{-0.01}$ ($0.07^{+0.01}_{-0.01}$) & $0.94^{+0.07}_{-0.09}$ ($0.72^{+0.03}_{-0.09}$) \\
\end{tabular}
\caption{This table summarizes the ejecta parameters and the 90\% confidence intervals (CI) for the posteriors obtained from the each kilonova morphological model when the luminosity distance is pinned to 40.7Mpc (and restricted inclination angle to the range found in \cite{Finstad_2018} in parentheses). \\}
\label{tab:newposteriorstats}
\end{table*}

\begin{figure*}
    \centering
    \includegraphics[width=\textwidth]{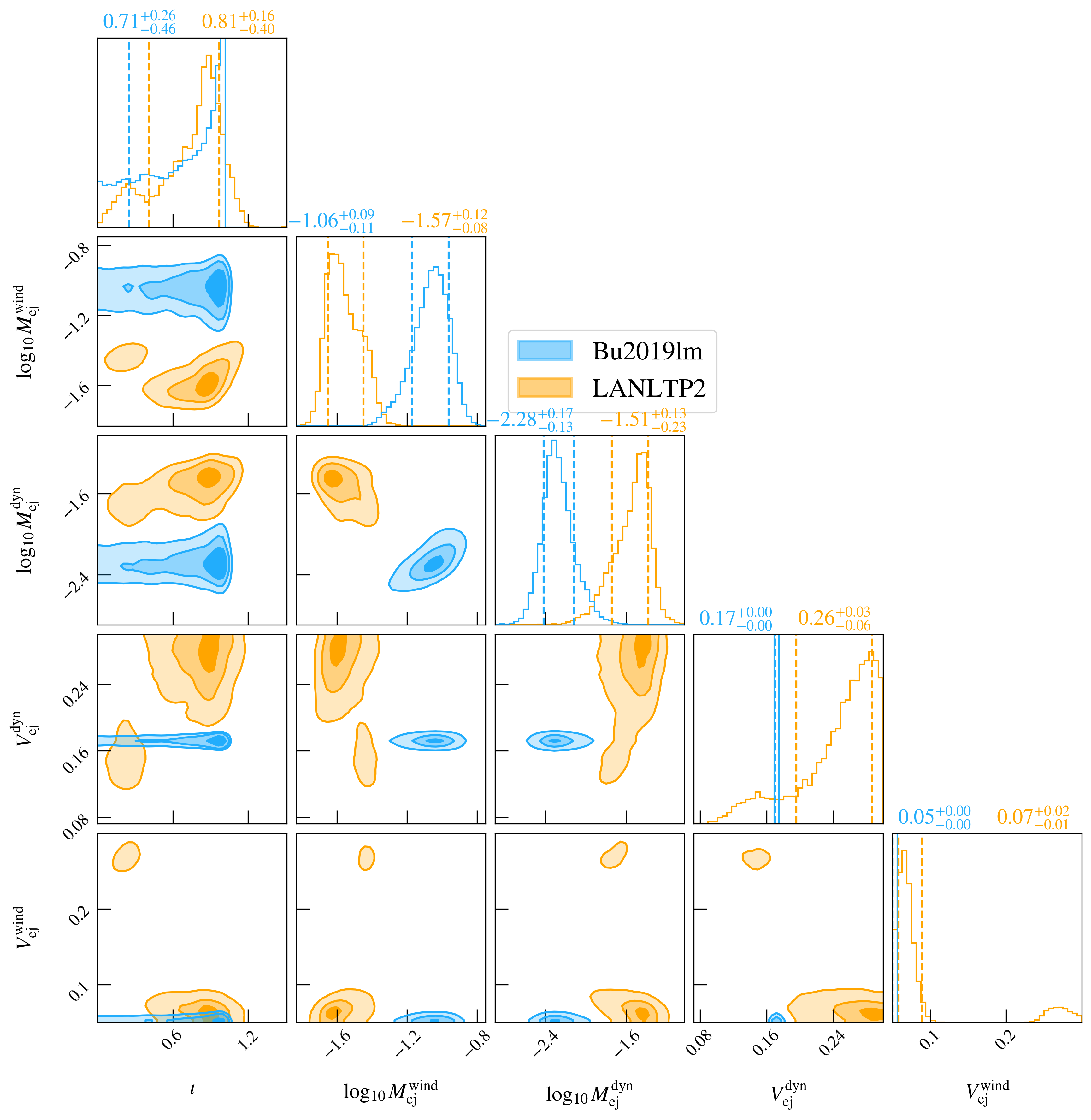}
    \caption{The comparison between the POSSIS-Bu2019 and the SuperNu-TP2 models discussed above. We note that the wind velocity, inclination angle, and dynamical mass ejecta all agree. However the Bu2019 model largely overestimates the wind mass ejecta (0.245 $M_\odot$). The Bu2019 model does not vary the ejecta velocities so plotted are the average values of the velocities used to expand the materials in the POSSIS simulations \citep[See Figure~1 in][]{Bulla2023}.}
    \label{fig:Bu_LANL_comp}
\end{figure*}

We also compare the analysis using our models to the next most recent models from \cite{bulla19}.
Plotted together in Figure~\ref{fig:Bu_LANL_comp} are the posteriors using the 2019 models from POSSIS and the TP2 model.
The TP2 model is favored over the POSSIS model for the same reason TP2 is preferred to the TS2 model above: the inferred values for the wind mass and velocity.
Wind mass is said to be ejected from the accretion disk surrounding the merger remnant and studies show $\approx30-40\%$ of the disk is ejected \citep{Sprouse_2024}.
Also low wind velocities naturally lead to a smaller fraction of wind mass ejection.
The value inferred from the POSSIS model points to an accretion disc of $0.2-0.3 M_{\odot}$ whereas the SuperNu model leads to an accretion disc of $0.06-0.09 M_{\odot}$.
From numerical relativity simulations, we know that accretion disks around the remnant BH should be of the order of $10^{-1} - 10^{-3} M_{\odot}$ (See Figure~S7 in \cite{Dietrich_2020}), so smaller value is more plausible.

%%%%%%%%%%%%%%%%%%%%%%%%%%%%%%%%%%%%%%%%%%%%%%%%%%%%%%%%%%%%%%%%%%%%%%%%%%%%%%
\section{Conclusion}
\label{sec:conclusion}
We propose that the ejecta from the merger event AT2017gfo/GW170817 follows a distribution similar to the TP2 morphology explained in this paper.
Using a neural network to map ejecta parameters to light curves, we found a maximal likelihood set of parameters for the optical observation of AT2017gfo listed in Table~\ref{tab:newposteriorstats}.
These parameters match other interpretations of AT2017gfo best for the TP2 morphology including a toroidal lanthanide-rich region in the equatorial plane and a `peanut' shaped distribution of lanthanide-free material in the polar region.
Specifically, the TP2 morphology matches the inferred inclination angle of \cite{hotokezaka2018hubble} best and predicts a more reasonable wind mass than the POSSIS model.
We showed that the emulators in NMMA for the new SuperNu kilonova models behave well in several validation tests.
Emulators accurately reproduce light curves when input parameters are in between grid points of the training set.
They also make up distinct morphological classes inside NMMA, meaning one morphology cannot be interpreted as another if observational data is abundant and accurate enough.
The emulators also show great synergy with the Bayesian inference network by producing light curves from input parameters that can correctly be interpreted and those parameters sufficiently recovered.
Finally, we showed that with some planned ToO strategies of the Vera Rubin Observatory and JWST, it is possible to distinguish ejecta morphologies.
With only a handful of new observations with these instruments, it is likely we can learn more about the morphology of each kilonova or about trends in ejecta morphology as a whole.
New information about kilonova morphology will help inform better modeling and elucidate the $r$-process happening in the NSMs.

%%%%%%%%%%%%%%%%%%%%%%%%%%%%%%%%%%%%%%%%%%%%%%%%%%%%%%%%%%%%%%%%%%%%%%%%%%%%%%
\section{Acknowledgements}
\label{sec:ack}
We thank Chris Fryer, Ryan Wollaeger, Marko Ristic, Kelsey Lund, Daniel Warshofsky, Andrew Toivenen, Eve Chase, and Ingo Tews for useful discussions.
B.K., S.D., and O.K. were supported by the Los Alamos National Laboratory (LANL) through its Center for Space and Earth Science (CSES). CSES is funded by LANL’s Laboratory Directed Research and Development (LDRD) program under project number 20240477CR-SES.
S.D. was also supported by the Laboratory Directed Research and Development program of Los Alamos National Laboratory under project number 20230315ER.
M.W.C. acknowledges support from the National Science Foundation with grant numbers PHY-2117997, PHY-2308862 and PHY-2409481.
P.T.H.P. is supported by the research program of the Netherlands Organization for Scientific Research (NWO) under grant number VI.Veni.232.021. 
This work used resources provided by the Los Alamos National Laboratory (LANL) Institutional Computing Program. 
LANL is operated by Triad National Security, LLC, for the National Nuclear Security Administration of the U.S.DOE  (Contract No. 89233218CNA000001). 
This work is authorized for unlimited release under LA-UR-25-24747 

% \bibliographystyle{aasjournal}
% \bibliography{refs}

\begin{thebibliography}{}
\expandafter\ifx\csname natexlab\endcsname\relax\def\natexlab#1{#1}\fi
\providecommand{\url}[1]{\href{#1}{#1}}
\providecommand{\dodoi}[1]{doi:~\href{http://doi.org/#1}{\nolinkurl{#1}}}
\providecommand{\doeprint}[1]{\href{http://ascl.net/#1}{\nolinkurl{http://ascl.net/#1}}}
\providecommand{\doarXiv}[1]{\href{https://arxiv.org/abs/#1}{\nolinkurl{https://arxiv.org/abs/#1}}}

\bibitem[{{Abbott} {et~al.}(2017){Abbott}, {Abbott}, {Abbott}, {Acernese},
  {Ackley}, {Adams}, {Adams}, {Addesso}, {Adhikari}, {Adya}, \&
  et~al.}]{abbott17a}
{Abbott}, B.~P., {Abbott}, R., {Abbott}, T.~D., {et~al.} 2017, \apjl, 848, L12,
  \dodoi{10.3847/2041-8213/aa91c9}

\bibitem[{{Alexander} {et~al.}(2017){Alexander}, {Berger}, {Fong}, {Williams},
  {Guidorzi}, {Margutti}, {Metzger}, {Annis}, {Blanchard}, {Brout}, {Brown},
  {Chen}, {Chornock}, {Cowperthwaite}, {Drout}, {Eftekhari}, {Frieman}, {Holz},
  {Nicholl}, {Rest}, {Sako}, {Soares-Santos}, \& {Villar}}]{alexander17}
{Alexander}, K.~D., {Berger}, E., {Fong}, W., {et~al.} 2017, \apjl, 848, L21,
  \dodoi{10.3847/2041-8213/aa905d}

\bibitem[{Anand {et~al.}(2020)Anand, Coughlin, Kasliwal, Bulla, Ahumada,
  Sagués~Carracedo, Almualla, Andreoni, Stein, Foucart, Singer, Sollerman,
  Bellm, Bolin, Caballero-García, Castro-Tirado, Cenko, De, Dekany, Duev,
  Feeney, Fremling, Goldstein, Golkhou, Graham, Guessoum, Hankins, Hu, Kong,
  Kool, Kulkarni, Kumar, Laher, Masci, Mróz, Nissanke, Porter, Reusch, Riddle,
  Rosnet, Rusholme, Serabyn, Sánchez-Ramírez, Rigault, Shupe, Smith,
  Soumagnac, Walters, \& Valeev}]{anand20}
Anand, S., Coughlin, M.~W., Kasliwal, M.~M., {et~al.} 2020, Nature Astronomy,
  5, 46–53, \dodoi{10.1038/s41550-020-1183-3}

\bibitem[{{Andreoni} \& {Margutti}(2024)}]{rubin24}
{Andreoni}, I., \& {Margutti}, R. 2024, Rubin ToO 2024: Envisioning the Vera C.
  Rubin Observatory LSST Target of Opportunity program, Tech. rep., Vera C.
  Rubin Observatory

\bibitem[{Andreoni {et~al.}(2017)Andreoni, Ackley, Cooke, Acharyya, Allison,
  Anderson, Ashley, Baade, Bailes, Bannister, Beardsley, Bessell, Bian, Bland,
  Boer, Booler, Brandeker, Brown, Buckley, Chang, Coward, Crawford, Crisp,
  Crosse, Cucchiara, Cupák, de~Gois, Deller, Devillepoix, Dobie, Elmer,
  Emrich, Farah, Farrell, Franzen, Gaensler, Galloway, Gendre, Giblin, Goobar,
  Green, Hancock, Hartig, Howell, Horsley, Hotan, Howie, Hu, Hu, James,
  Johnston, Johnston-Hollitt, Kaplan, Kasliwal, Keane, Kenney, Klotz, Lau,
  Laugier, Lenc, Li, Liang, Lidman, Luvaul, Lynch, Ma, Macpherson, Mao,
  McClelland, McCully, Möller, Morales, Morris, Murphy, Noysena, Onken,
  Orange, Osłowski, Pallot, Paxman, Potter, Pritchard, Raja, Ridden-Harper,
  Romero-Colmenero, Sadler, Sansom, Scalzo, Schmidt, Scott, Seghouani, Shang,
  Shannon, Shao, Shara, Sharp, Sokolowski, Sollerman, Staff, Steele, Sun,
  Suntzeff, Tao, Tingay, Towner, Thierry, Trott, Tucker, Väisänen, Krishnan,
  Walker, Wang, Wang, Wayth, Whiting, Williams, Williams, Wolf, Wu, Wu, Yang,
  Yuan, Zhang, Zhou, \& Zovaro}]{Andreoni_2017}
Andreoni, I., Ackley, K., Cooke, J., {et~al.} 2017, Publications of the
  Astronomical Society of Australia, 34, \dodoi{10.1017/pasa.2017.65}

\bibitem[{{Arcavi} {et~al.}(2017){Arcavi}, {Hosseinzadeh}, {Howell}, {McCully},
  {Poznanski}, {Kasen}, {Barnes}, {Zaltzman}, {Vasylyev}, {Maoz}, \&
  {Valenti}}]{arcavi17n}
{Arcavi}, I., {Hosseinzadeh}, G., {Howell}, D.~A., {et~al.} 2017, \nat, 551,
  64, \dodoi{10.1038/nature24291}

\bibitem[{Barna {et~al.}(2024)Barna, Reed, Andreoni, Coughlin, Dietrich, Groom,
  du~Laz, Pang, Purdum, \& Rusholme}]{barna2024}
Barna, T., Reed, B., Andreoni, I., {et~al.} 2024, An Online Framework for
  Fitting Fast Transient Lightcurves.
\newblock \doarXiv{2404.17515}

\bibitem[{Buchner {et~al.}(2014)Buchner, Georgakakis, Nandra, Hsu, Rangel,
  Brightman, Merloni, Salvato, Donley, \& Kocevski}]{Buchner:2014nha}
Buchner, J., Georgakakis, A., Nandra, K., {et~al.} 2014, Astron. Astrophys.,
  564, A125, \dodoi{10.1051/0004-6361/201322971}

\bibitem[{Bulla(2019)}]{bulla19}
Bulla, M. 2019, Monthly Notices of the Royal Astronomical Society, 489, 5037,
  \dodoi{10.1093/mnras/stz2495}

\bibitem[{{Bulla}(2023)}]{Bulla2023}
{Bulla}, M. 2023, 363, 241, \dodoi{10.1017/S1743921322000424}

\bibitem[{Burns(2020)}]{Burns_2020}
Burns, E. 2020, Living Reviews in Relativity, 23,
  \dodoi{10.1007/s41114-020-00028-7}

\bibitem[{Chambers {et~al.}(2019)Chambers, Magnier, Metcalfe, Flewelling,
  Huber, Waters, Denneau, Draper, Farrow, Finkbeiner, Holmberg, Koppenhoefer,
  Price, Rest, Saglia, Schlafly, Smartt, Sweeney, Wainscoat, Burgett, Chastel,
  Grav, Heasley, Hodapp, Jedicke, Kaiser, Kudritzki, Luppino, Lupton, Monet,
  Morgan, Onaka, Shiao, Stubbs, Tonry, White, Bañados, Bell, Bender, Bernard,
  Boegner, Boffi, Botticella, Calamida, Casertano, Chen, Chen, Cole, Deacon,
  Frenk, Fitzsimmons, Gezari, Gibbs, Goessl, Goggia, Gourgue, Goldman, Grant,
  Grebel, Hambly, Hasinger, Heavens, Heckman, Henderson, Henning, Holman, Hopp,
  Ip, Isani, Jackson, Keyes, Koekemoer, Kotak, Le, Liska, Long, Lucey, Liu,
  Martin, Masci, McLean, Mindel, Misra, Morganson, Murphy, Obaika, Narayan,
  Nieto-Santisteban, Norberg, Peacock, Pier, Postman, Primak, Rae, Rai, Riess,
  Riffeser, Rix, Röser, Russel, Rutz, Schilbach, Schultz, Scolnic, Strolger,
  Szalay, Seitz, Small, Smith, Soderblom, Taylor, Thomson, Taylor, Thakar,
  Thiel, Thilker, Unger, Urata, Valenti, Wagner, Walder, Walter, Watters,
  Werner, Wood-Vasey, \& Wyse}]{chambers2019panstarrs1surveys}
Chambers, K.~C., Magnier, E.~A., Metcalfe, N., {et~al.} 2019, The Pan-STARRS1
  Surveys.
\newblock \doarXiv{1612.05560}

\bibitem[{{Chornock} {et~al.}(2017){Chornock}, {Berger}, {Kasen},
  {Cowperthwaite}, {Nicholl}, {Villar}, {Alexander}, {Blanchard}, {Eftekhari},
  {Fong}, {Margutti}, {Williams}, {Annis}, {Brout}, {Brown}, {Chen}, {Drout},
  {Farr}, {Foley}, {Frieman}, {Fryer}, {Herner}, {Holz}, {Kessler}, {Matheson},
  {Metzger}, {Quataert}, {Rest}, {Sako}, {Scolnic}, {Smith}, \&
  {Soares-Santos}}]{chornock17}
{Chornock}, R., {Berger}, E., {Kasen}, D., {et~al.} 2017, \apjl, 848, L19,
  \dodoi{10.3847/2041-8213/aa905c}

\bibitem[{Coughlin {et~al.}(2019)Coughlin, Ahumada, Anand, De, Hankins,
  Kasliwal, Singer, Bellm, Andreoni, Cenko, Cooke, Copperwheat, Dugas, Jencson,
  Perley, Yu, Bhalerao, Kumar, Bloom, Anupama, Ashley, Bagdasaryan, Biswas,
  Buckley, Burdge, Cook, Cromer, Cunningham, D’Aì, Dekany, Delacroix,
  Dichiara, Duev, Dutta, Feeney, Frederick, Gatkine, Ghosh, Goldstein, Golkhou,
  Goobar, Graham, Hanayama, Horiuchi, Hung, Jha, Kong, Giomi, Kaplan,
  Karambelkar, Kowalski, Kulkarni, Kupfer, Masci, Mazzali, Moore, Mogotsi,
  Neill, Ngeow, Martínez-Palomera, Parola, Pavana, Ofek, Patil, Riddle,
  Rigault, Rusholme, Serabyn, Shupe, Sharma, Singh, Sollerman, Soon, Staats,
  Taggart, Tan, Travouillon, Troja, Waratkar, \& Yatsu}]{coughlin19}
Coughlin, M.~W., Ahumada, T., Anand, S., {et~al.} 2019, The Astrophysical
  Journal Letters, 885, L19, \dodoi{10.3847/2041-8213/ab4ad8}

\bibitem[{{Coulter} {et~al.}(2017){Coulter}, {Foley}, {Kilpatrick}, {Drout},
  {Piro}, {Shappee}, {Siebert}, {Simon}, {Ulloa}, {Kasen}, {Madore},
  {Murguia-Berthier}, {Pan}, {Prochaska}, {Ramirez-Ruiz}, {Rest}, \&
  {Rojas-Bravo}}]{coulter17}
{Coulter}, D.~A., {Foley}, R.~J., {Kilpatrick}, C.~D., {et~al.} 2017, Science,
  \dodoi{doi:10.1126/science.aap9811}

\bibitem[{{Cowan} {et~al.}(2021){Cowan}, {Sneden}, {Lawler}, {Aprahamian},
  {Wiescher}, {Langanke}, {Mart{\'\i}nez-Pinedo}, \& {Thielemann}}]{cowan21}
{Cowan}, J.~J., {Sneden}, C., {Lawler}, J.~E., {et~al.} 2021, Reviews of Modern
  Physics, 93, 015002, \dodoi{10.1103/RevModPhys.93.015002}

\bibitem[{{Cowperthwaite} {et~al.}(2017){Cowperthwaite}, {Berger}, {Villar},
  {Metzger}, {Nicholl}, {Chornock}, {Blanchard}, {Fong}, {Margutti}, \&
  {Soares-Santos}}]{cowperthwaite17}
{Cowperthwaite}, P.~S., {Berger}, E., {Villar}, V.~A., {et~al.} 2017, \apj,
  848, L17, \dodoi{10.3847/2041-8213/aa8fc7}

\bibitem[{{D{\'{\i}}az} {et~al.}(2017){D{\'{\i}}az}, {Macri}, {Garcia Lambas},
  {Mendes de Oliveira}, {Nilo Castell{\'o}n}, {Ribeiro}, {S{\'a}nchez},
  {Schoenell}, {Abramo}, {Akras}, {Alcaniz}, {Artola}, {Beroiz}, {Bonoli},
  {Cabral}, {Camuccio}, {Castillo}, {Chavushyan}, {Coelho}, {Colazo},
  {Costa-Duarte}, {Cuevas Larenas}, {DePoy}, {Dom{\'{\i}}nguez Romero},
  {Dultzin}, {Fern{\'a}ndez}, {Garc{\'{\i}}a}, {Girardini}, {Gon{\c c}alves},
  {Gon{\c c}alves}, {Gurovich}, {Jim{\'e}nez-Teja}, {Kanaan}, {Lares}, {Lopes
  de Oliveira}, {L{\'o}pez-Cruz}, {Marshall}, {Melia}, {Molino}, {Padilla},
  {Pe{\~n}uela}, {Placco}, {Qui{\~n}ones}, {Ram{\'{\i}}rez Rivera}, {Renzi},
  {Riguccini}, {R{\'{\i}}os-L{\'o}pez}, {Rodriguez}, {Sampedro}, {Schneiter},
  {Sodr{\'e}}, {Starck}, {Torres-Flores}, {Tornatore}, \&
  {Zadro{\.z}ny}}]{diaz17}
{D{\'{\i}}az}, M.~C., {Macri}, L.~M., {Garcia Lambas}, D., {et~al.} 2017,
  \apjl, 848, L29, \dodoi{10.3847/2041-8213/aa9060}

\bibitem[{Dicken {et~al.}(2024)Dicken, Marín, Shivaei, Guillard, Libralato,
  Glasse, Gordon, Cossou, Kavanagh, Temim, Flagey, Klaassen, Rieke, Wright,
  Alberts, Azzollini, Álvarez Márquez, Bouchet, Bright, Cracraft, Coulais,
  Detre, Engesser, Fox, Gaspar, Gastaud, Glauser, Hines, Kendrew, Labiano,
  Lagage, Lee, Law, Morrison, Noriega-Crespo, Jones, Patapis, Scheithauer,
  Sloan, \& Tamas}]{Dicken_2024}
Dicken, D., Marín, M.~G., Shivaei, I., {et~al.} 2024, Astronomy \&
  Astrophysics, 689, A5, \dodoi{10.1051/0004-6361/202449451}

\bibitem[{Dietrich {et~al.}(2020)Dietrich, Coughlin, Pang, Bulla, Heinzel,
  Issa, Tews, \& Antier}]{Dietrich_2020}
Dietrich, T., Coughlin, M.~W., Pang, P. T.~H., {et~al.} 2020, Science, 370,
  1450–1453, \dodoi{10.1126/science.abb4317}

\bibitem[{Dietrich \& Ujevic(2017)}]{Dietrich_2017}
Dietrich, T., \& Ujevic, M. 2017, Classical and Quantum Gravity, 34, 105014,
  \dodoi{10.1088/1361-6382/aa6bb0}

\bibitem[{{Drout} {et~al.}(2017){Drout}, {Piro}, {Shappee}, {Kilpatrick},
  {Simon}, {Contreras}, {Coulter}, {Foley}, {Siebert}, {Morrell}, {Boutsia},
  {Di Mille}, {Holoien}, {Kasen}, {Kollmeier}, {Madore}, {Monson},
  {Murguia-Berthier}, {Pan}, {Prochaska}, {Ramirez-Ruiz}, {Rest}, {Adams},
  {Alatalo}, {Ba{\~n}ados}, {Baughman}, {Beers}, {Bernstein}, {Bitsakis},
  {Campillay}, {Hansen}, {Higgs}, {Ji}, {Maravelias}, {Marshall}, {Moni Bidin},
  {Prieto}, {Rasmussen}, {Rojas-Bravo}, {Strom}, {Ulloa},
  {Vargas-Gonz{\'a}lez}, {Wan}, \& {Whitten}}]{drout17}
{Drout}, M.~R., {Piro}, A.~L., {Shappee}, B.~J., {et~al.} 2017, Science, in
  press, available via doi:10.1126/science.aaq0049,
  \dodoi{10.1126/science.aaq0049}

\bibitem[{{Evans} {et~al.}(2017){Evans}, {Cenko}, {Kennea}, {Emery}, {Kuin},
  {Korobkin}, {Wollaeger}, {Fryer}, {Madsen}, {Harrison}, {Xu}, {Nakar},
  {Hotokezaka}, {Lien}, {Campana}, {Oates}, {Troja}, {Breeveld}, {Marshall},
  {Barthelmy}, {Beardmore}, {Burrows}, {Cusumano}, {D'Ai}, {D'Avanzo},
  {D'Elia}, {de Pasquale}, {Even}, {Fontes}, {Forster}, {Garcia}, {Giommi},
  {Grefenstette}, {Gronwall}, {Hartmann}, {Heida}, {Hungerford}, {Kasliwal},
  {Krimm}, {Levan}, {Malesani}, {Melandri}, {Miyasaka}, {Nousek}, {O'Brien},
  {Osborne}, {Pagani}, {Page}, {Palmer}, {Perri}, {Pike}, {Racusin}, {Rosswog},
  {Siegel}, {Sakamoto}, {Sbarufatti}, {Tagliaferri}, {Tanvir}, \&
  {Tohuvavohu}}]{evans17}
{Evans}, P.~A., {Cenko}, S.~B., {Kennea}, J.~A., {et~al.} 2017, Science,
  \dodoi{doi.org/10.1126/science.aap9580}

\bibitem[{{Fern{\'a}ndez} {et~al.}(2015){Fern{\'a}ndez}, {Quataert}, {Schwab},
  {Kasen}, \& {Rosswog}}]{fernandez15}
{Fern{\'a}ndez}, R., {Quataert}, E., {Schwab}, J., {Kasen}, D., \& {Rosswog},
  S. 2015, \mnras, 449, 390, \dodoi{10.1093/mnras/stv238}

\bibitem[{Feroz {et~al.}(2009)Feroz, Hobson, \& Bridges}]{Feroz:2008xx}
Feroz, F., Hobson, M.~P., \& Bridges, M. 2009, Mon. Not. Roy. Astron. Soc.,
  398, 1601, \dodoi{10.1111/j.1365-2966.2009.14548.x}

\bibitem[{Finstad {et~al.}(2018)Finstad, De, Brown, Berger, \&
  Biwer}]{Finstad_2018}
Finstad, D., De, S., Brown, D.~A., Berger, E., \& Biwer, C.~M. 2018, The
  Astrophysical Journal Letters, 860, L2, \dodoi{10.3847/2041-8213/aac6c1}

\bibitem[{{Fontes} {et~al.}(2015){Fontes}, {Fryer}, {Hungerford}, {Hakel},
  {Colgan}, {Kilcrease}, \& {Sherrill}}]{fontes15a}
{Fontes}, C.~J., {Fryer}, C.~L., {Hungerford}, A.~L., {et~al.} 2015, {High
  Energy Density Physics}, 16, 53,
  \dodoi{http://dx.doi.org/10.1016/j.hedp.2015.06.002}

\bibitem[{{Fontes} {et~al.}(2020){Fontes}, {Fryer}, {Hungerford}, {Wollaeger},
  \& {Korobkin}}]{fontes20}
{Fontes}, C.~J., {Fryer}, C.~L., {Hungerford}, A.~L., {Wollaeger}, R.~T., \&
  {Korobkin}, O. 2020, arXiv e-prints, arXiv:1904.08781.
\newblock \doarXiv{1904.08781}

\bibitem[{{Freiburghaus} {et~al.}(1999){Freiburghaus}, {Rosswog}, \&
  {Thielemann}}]{freiburghaus99}
{Freiburghaus}, C., {Rosswog}, S., \& {Thielemann}, F.~K. 1999, \apj, 525,
  L121, \dodoi{10.1086/312343}

\bibitem[{{Grossman} {et~al.}(2014){Grossman}, {Korobkin}, {Rosswog}, \&
  {Piran}}]{grossman14}
{Grossman}, D., {Korobkin}, O., {Rosswog}, S., \& {Piran}, T. 2014, \mnras,
  439, 757, \dodoi{10.1093/mnras/stt2503}

\bibitem[{Hjorth {et~al.}(2017)Hjorth, Levan, Tanvir, Lyman, Wojtak, Schrøder,
  Mandel, Gall, \& Bruun}]{Hjorth_2017}
Hjorth, J., Levan, A.~J., Tanvir, N.~R., {et~al.} 2017, The Astrophysical
  Journal, 848, L31, \dodoi{10.3847/2041-8213/aa9110}

\bibitem[{Hotokezaka {et~al.}(2018)Hotokezaka, Nakar, Gottlieb, Nissanke,
  Masuda, Hallinan, Mooley, \& Deller}]{hotokezaka2018hubble}
Hotokezaka, K., Nakar, E., Gottlieb, O., {et~al.} 2018.
\newblock \doarXiv{1806.10596}

\bibitem[{Hotokezaka \& Piran(2015)}]{Hotokezaka_2015}
Hotokezaka, K., \& Piran, T. 2015, Monthly Notices of the Royal Astronomical
  Society, 450, 1430–1440, \dodoi{10.1093/mnras/stv620}

\bibitem[{{Hu} {et~al.}(2017){Hu}, {Wu}, {Andreoni}, {Ashley}, {Cooke}, {Cui},
  {Du}, {Dai}, {Gu}, {Hu}, {Lu}, {Li}, {Li}, {Liang}, {Liu}, {Ma}, {Shang},
  {Sun}, {Suntzeff}, {Tao}, {Uddin}, {Wang}, {Wang}, {Wen}, {Xiao}, {Xu},
  {Yang}, {Yang}, {Yuan}, {Zhou}, {Zhang}, {Zhou}, \& {Zhu}}]{hul17}
{Hu}, L., {Wu}, X., {Andreoni}, I., {et~al.} 2017, ArXiv e-prints.
\newblock \doarXiv{1710.05462}

\bibitem[{Jarrett {et~al.}(2000)Jarrett, Chester, Cutri, Schneider, Skrutskie,
  \& Huchra}]{Jarrett_2000}
Jarrett, T.~H., Chester, T., Cutri, R., {et~al.} 2000, The Astronomical
  Journal, 119, 2498–2531, \dodoi{10.1086/301330}

\bibitem[{{Kasliwal} {et~al.}(2017{\natexlab{a}}){Kasliwal}, {Korobkin}, {Lau},
  {Wollaeger}, \& {Fryer}}]{kasliwal17a}
{Kasliwal}, M.~M., {Korobkin}, O., {Lau}, R.~M., {Wollaeger}, R., \& {Fryer},
  C.~L. 2017{\natexlab{a}}, \apj, 843, L34, \dodoi{10.3847/2041-8213/aa799d}

\bibitem[{{Kasliwal} {et~al.}(2017{\natexlab{b}}){Kasliwal}, {Nakar}, {Singer},
  {Kaplan}, {Cook}, {Van Sistine}, {Lau}, {Fremling}, {Gottlieb}, {Jencson},
  {Adams}, {Feindt}, {Hotokezaka}, {Ghosh}, {Perley}, {Yu}, {Piran}, {Allison},
  {Anupama}, {Balasubramanian}, {Bannister}, {Bally}, {Barnes}, {Barway},
  {Bellm}, {Bhalerao}, {Bhattacharya}, {Blagorodnova}, {Bloom}, {Brady},
  {Cannella}, {Chatterjee}, {Cenko}, {Cobb}, {Copperwheat}, {Corsi}, {De},
  {Dobie}, {Emery}, {Evans}, {Fox}, {Frail}, {Frohmaier}, {Goobar}, {Hallinan},
  {Harrison}, {Helou}, {Hinderer}, {Ho}, {Horesh}, {Ip}, {Itoh}, {Kasen},
  {Kim}, {Kuin}, {Kupfer}, {Lynch}, {Madsen}, {Mazzali}, {Miller}, {Mooley},
  {Murphy}, {Ngeow}, {Nichols}, {Nissanke}, {Nugent}, {Ofek}, {Qi}, {Quimby},
  {Rosswog}, {Rusu}, {Sadler}, {Schmidt}, {Sollerman}, {Steele}, {Williamson},
  {Xu}, {Yan}, {Yatsu}, {Zhang}, \& {Zhao}}]{kasliwal17b}
{Kasliwal}, M.~M., {Nakar}, E., {Singer}, L.~P., {et~al.} 2017{\natexlab{b}},
  Science in press, available via doi:10.1126/science.aap9455,
  \dodoi{10.1126/science.aap9455}

\bibitem[{Kiendrebeogo {et~al.}(2023)Kiendrebeogo, Farah, Foley, Gray, Kunert,
  Puecher, Toivonen, VandenBerg, Anand, Ahumada, Karambelkar, Coughlin,
  Dietrich, Kam, Pang, Singer, \& Sravan}]{Kiendrebeogo_2023}
Kiendrebeogo, R.~W., Farah, A.~M., Foley, E.~M., {et~al.} 2023, The
  Astrophysical Journal, 958, 158, \dodoi{10.3847/1538-4357/acfcb1}

\bibitem[{{Korobkin} {et~al.}(2012){Korobkin}, {Rosswog}, {Arcones}, \&
  {Winteler}}]{korobkin12}
{Korobkin}, O., {Rosswog}, S., {Arcones}, A., \& {Winteler}, C. 2012, \mnras,
  426, 1940, \dodoi{10.1111/j.1365-2966.2012.21859.x}

\bibitem[{Korobkin {et~al.}(2021)Korobkin, Wollaeger, Fryer, Hungerford,
  Rosswog, Fontes, Mumpower, Chase, Even, Miller, Misch, \&
  Lippuner}]{korobkin21}
Korobkin, O., Wollaeger, R.~T., Fryer, C.~L., {et~al.} 2021, The Astrophysical
  Journal, 910, 116, \dodoi{10.3847/1538-4357/abe1b5}

\bibitem[{{Lipunov} {et~al.}(2017){Lipunov}, {Kornilov}, {Gorbovskoy},
  {Lipunova}, {Vlasenko}, {Panchenko}, {Tyurina}, \& {Grinshpun}}]{lipunov17o}
{Lipunov}, V., {Kornilov}, V., {Gorbovskoy}, E., {et~al.} 2017, ArXiv e-prints.
\newblock \doarXiv{1710.05911}

\bibitem[{{Margutti} {et~al.}(2017){Margutti}, {Berger}, {Fong}, {Guidorzi},
  {Alexander}, {Metzger}, {Blanchard}, {Cowperthwaite}, {Chornock},
  {Eftekhari}, {Nicholl}, {Villar}, {Williams}, {Annis}, {Brown}, {Chen},
  {Doctor}, {Frieman}, {Holz}, {Sako}, \& {Soares-Santos}}]{margutti17}
{Margutti}, R., {Berger}, E., {Fong}, W., {et~al.} 2017, \apjl, 848, L20,
  \dodoi{10.3847/2041-8213/aa9057}

\bibitem[{{Metzger}(2017)}]{metzger17y}
{Metzger}, B.~D. 2017, Living Reviews in Relativity, 20, 3,
  \dodoi{10.1007/s41114-017-0006-z}

\bibitem[{{Metzger} {et~al.}(2010){Metzger}, {Mart{\'\i}nez-Pinedo}, {Darbha},
  {Quataert}, {Arcones}, {Kasen}, {Thomas}, {Nugent}, {Panov}, \&
  {Zinner}}]{metzger10}
{Metzger}, B.~D., {Mart{\'\i}nez-Pinedo}, G., {Darbha}, S., {et~al.} 2010,
  \mnras, 406, 2650, \dodoi{10.1111/j.1365-2966.2010.16864.x}

\bibitem[{{Miller} {et~al.}(2019{\natexlab{a}}){Miller}, {Ryan}, \&
  {Dolence}}]{miller19a}
{Miller}, J.~M., {Ryan}, B.~R., \& {Dolence}, J.~C. 2019{\natexlab{a}}, \apjs,
  241, 30, \dodoi{10.3847/1538-4365/ab09fc}

\bibitem[{{Miller} {et~al.}(2019{\natexlab{b}}){Miller}, {Ryan}, {Dolence},
  {Burrows}, {Fontes}, {Fryer}, {Korobkin}, {Lippuner}, {Mumpower}, \&
  {Wollaeger}}]{miller19b}
{Miller}, J.~M., {Ryan}, B.~R., {Dolence}, J.~C., {et~al.} 2019{\natexlab{b}},
  arXiv e-prints, arXiv:1905.07477.
\newblock \doarXiv{1905.07477}

\bibitem[{{Pang} {et~al.}(2022){Pang}, {Dietrich}, {Coughlin}, {Bulla}, {Tews},
  {Almualla}, {Barna}, {Kiendrebeogo}, {Kunert}, {Mansingh}, {Reed}, {Sravan},
  {Toivonen}, {Antier}, {VandenBerg}, {Heinzel}, {Nedora}, {Salehi}, {Sharma},
  {Somasundaram}, \& {Van Den Broeck}}]{pang22}
{Pang}, P. T.~H., {Dietrich}, T., {Coughlin}, M.~W., {et~al.} 2022, arXiv
  e-prints, arXiv:2205.08513, \dodoi{10.48550/arXiv.2205.08513}

\bibitem[{{Pian} {et~al.}(2017){Pian}, {D'Avanzo}, {Benetti}, {Branchesi},
  {Brocato}, {Campana}, {Cappellaro}, {Covino}, {D'Elia}, {Fynbo}, {Getman},
  {Ghirlanda}, {Ghisellini}, {Grado}, {Greco}, {Hjorth}, {Kouveliotou},
  {Levan}, {Limatola}, {Malesani}, {Mazzali}, {Melandri}, {M{\o}ller},
  {Nicastro}, {Palazzi}, {Piranomonte}, {Rossi}, {Salafia}, {Selsing},
  {Stratta}, {Tanaka}, {Tanvir}, {Tomasella}, {Watson}, {Yang}, {Amati},
  {Antonelli}, {Ascenzi}, {Bernardini}, {Bo{\"e}r}, {Bufano}, {Bulgarelli},
  {Capaccioli}, {Casella}, {Castro-Tirado}, {Chassande-Mottin}, {Ciolfi},
  {Copperwheat}, {Dadina}, {De Cesare}, {di Paola}, {Fan}, {Gendre},
  {Giuffrida}, {Giunta}, {Hunt}, {Israel}, {Jin}, {Kasliwal}, {Klose}, {Lisi},
  {Longo}, {Maiorano}, {Mapelli}, {Masetti}, {Nava}, {Patricelli}, {Perley},
  {Pescalli}, {Piran}, {Possenti}, {Pulone}, {Razzano}, {Salvaterra},
  {Schipani}, {Spera}, {Stamerra}, {Stella}, {Tagliaferri}, {Testa}, {Troja},
  {Turatto}, {Vergani}, \& {Vergani}}]{pian17}
{Pian}, E., {D'Avanzo}, P., {Benetti}, S., {et~al.} 2017, \nat, 551, 67,
  \dodoi{10.1038/nature24298}

\bibitem[{{Radice} \& {Dai}(2019)}]{radice19}
{Radice}, D., \& {Dai}, L. 2019, European Physical Journal A, 55, 50,
  \dodoi{10.1140/epja/i2019-12716-4}

\bibitem[{{Ristic} {et~al.}(2023){Ristic}, {O'Shaughnessy}, {Villar},
  {Wollaeger}, {Korobkin}, {Fryer}, {Fontes}, \& {Kedia}}]{ristic23b}
{Ristic}, M., {O'Shaughnessy}, R., {Villar}, V.~A., {et~al.} 2023, arXiv
  e-prints, arXiv:2304.06699, \dodoi{10.48550/arXiv.2304.06699}

\bibitem[{{Ristic} {et~al.}(2022){Ristic}, {Champion}, {O'Shaughnessy},
  {Wollaeger}, {Korobkin}, {Chase}, {Fryer}, {Hungerford}, \&
  {Fontes}}]{ristic22}
{Ristic}, M., {Champion}, E., {O'Shaughnessy}, R., {et~al.} 2022, Physical
  Review Research, 4, 013046, \dodoi{10.1103/PhysRevResearch.4.013046}

\bibitem[{{Rosswog}(2013)}]{rosswog13}
{Rosswog}, S. 2013, \nat, 500, 535, \dodoi{10.1038/500535a}

\bibitem[{{Shappee} {et~al.}(2017){Shappee}, {Simon}, {Drout}, {Piro},
  {Morrell}, {Prieto}, {Kasen}, {Holoien}, {Kollmeier}, {Kelson}, {Coulter},
  {Foley}, {Kilpatrick}, {Siebert}, {Madore}, {Murguia-Berthier}, {Pan},
  {Prochaska}, {Ramirez-Ruiz}, {Rest}, {Adams}, {Alatalo}, {Ba{\~n}ados},
  {Baughman}, {Bernstein}, {Bitsakis}, {Boutsia}, {Bravo}, {Di Mille}, {Higgs},
  {Ji}, {Maravelias}, {Marshall}, {Placco}, {Prieto}, \& {Wan}}]{shappee17}
{Shappee}, B.~J., {Simon}, J.~D., {Drout}, M.~R., {et~al.} 2017, Science, 358,
  1574, \dodoi{10.1126/science.aaq0186}

\bibitem[{Skilling(2006)}]{skilling2006}
Skilling, J. 2006, Bayesian Anal., 1, 833, \dodoi{10.1214/06-BA127}

\bibitem[{{Smartt} {et~al.}(2017){Smartt}, {Chen}, {Jerkstrand}, {Coughlin},
  {Kankare}, {Sim}, {Fraser}, {Inserra}, {Maguire}, {Chambers}, {Huber},
  {Kr{\"u}hler}, {Leloudas}, {Magee}, {Shingles}, {Smith}, {Young}, {Tonry},
  {Kotak}, {Gal-Yam}, {Lyman}, {Homan}, {Agliozzo}, {Anderson}, {Angus},
  {Ashall}, {Barbarino}, {Bauer}, {Berton}, {Botticella}, {Bulla}, {Bulger},
  {Cannizzaro}, {Cano}, {Cartier}, {Cikota}, {Clark}, {De Cia}, {Della Valle},
  {Denneau}, {Dennefeld}, {Dessart}, {Dimitriadis}, {Elias-Rosa}, {Firth},
  {Flewelling}, {Fl{\"o}rs}, {Franckowiak}, {Frohmaier}, {Galbany},
  {Gonz{\'a}lez-Gait{\'a}n}, {Greiner}, {Gromadzki}, {Guelbenzu},
  {Guti{\'e}rrez}, {Hamanowicz}, {Hanlon}, {Harmanen}, {Heintz}, {Heinze},
  {Hernandez}, {Hodgkin}, {Hook}, {Izzo}, {James}, {Jonker}, {Kerzendorf},
  {Klose}, {Kostrzewa-Rutkowska}, {Kowalski}, {Kromer}, {Kuncarayakti},
  {Lawrence}, {Lowe}, {Magnier}, {Manulis}, {Martin-Carrillo}, {Mattila},
  {McBrien}, {M{\"u}ller}, {Nordin}, {ONeill}, {Onori}, {Palmerio},
  {Pastorello}, {Patat}, {Pignata}, {Podsiadlowski}, {Pumo}, {Prentice}, {Rau},
  {Razza}, {Rest}, {Reynolds}, {Roy}, {Ruiter}, {Rybicki}, {Salmon}, {Schady},
  {Schultz}, {Schweyer}, {Seitenzahl}, {Smith}, {Sollerman}, {Stalder},
  {Stubbs}, {Sullivan}, {Szegedi}, {Taddia}, {Taubenberger}, {Terreran}, {van
  Soelen}, {Vos}, {Wainscoat}, {Walton}, {Waters}, {Weiland}, {Willman},
  {Wiseman}, {Wright}, {Wyrzykowski}, \& {Yaron}}]{smartt17}
{Smartt}, S.~J., {Chen}, T.-W., {Jerkstrand}, A., {et~al.} 2017, \nat, 551, 75,
  \dodoi{10.1038/nature24303}

\bibitem[{Sneppen {et~al.}(2023)Sneppen, Watson, Bauswein, Just, Kotak, Nakar,
  Poznanski, \& Sim}]{Sneppen_2023}
Sneppen, A., Watson, D., Bauswein, A., {et~al.} 2023, Nature, 614, 436–439,
  \dodoi{10.1038/s41586-022-05616-x}

\bibitem[{{Soares-Santos} {et~al.}(2017){Soares-Santos}, {Holz}, {Annis},
  {Chornock}, {Herner}, {Berger}, {Brout}, {Chen}, {Kessler}, {Sako}, {Allam},
  {Tucker}, {Butler}, {Palmese}, {Doctor}, {Diehl}, {Frieman}, {Yanny}, {Lin},
  {Scolnic}, {Cowperthwaite}, {Neilsen}, {Marriner}, {Kuropatkin}, {Hartley},
  {Paz-Chinch{\'o}n}, {Alexander}, {Balbinot}, {Blanchard}, {Brown}, {Carlin},
  {Conselice}, {Cook}, {Drlica-Wagner}, {Drout}, {Durret}, {Eftekhari}, {Farr},
  {Finley}, {Foley}, {Fong}, {Fryer}, {Garc{\'{\i}}a-Bellido}, {Gill},
  {Gruendl}, {Hanna}, {Kasen}, {Li}, {Lopes}, {Louren{\c c}o}, {Margutti},
  {Marshall}, {Matheson}, {Medina}, {Metzger}, {Mu{\~n}oz}, {Muir}, {Nicholl},
  {Quataert}, {Rest}, {Sauseda}, {Schlegel}, {Secco}, {Sobreira}, {Stebbins},
  {Villar}, {Vivas}, {Walker}, {Wester}, {Williams}, {Zenteno}, {Zhang},
  {Abbott}, {Abdalla}, {Banerji}, {Bechtol}, {Benoit-L{\'e}vy}, {Bertin},
  {Brooks}, {Buckley-Geer}, {Burke}, {Carnero Rosell}, {Carrasco Kind},
  {Carretero}, {Castander}, {Crocce}, {Cunha}, {DAndrea}, {da Costa}, {Davis},
  {Desai}, {Dietrich}, {Doel}, {Eifler}, {Fernandez}, {Flaugher}, {Fosalba},
  {Gaztanaga}, {Gerdes}, {Giannantonio}, {Goldstein}, {Gruen}, {Gschwend},
  {Gutierrez}, {Honscheid}, {Jain}, {James}, {Jeltema}, {Johnson}, {Johnson},
  {Kent}, {Krause}, {Kron}, {Kuehn}, {Kuhlmann}, {Lahav}, {Lima}, {Maia},
  {March}, {McMahon}, {Menanteau}, {Miquel}, {Mohr}, {Nichol}, {Nord},
  {Ogando}, {Petravick}, {Plazas}, {Romer}, {Roodman}, {Rykoff}, {Sanchez},
  {Scarpine}, {Schubnell}, {Sevilla-Noarbe}, {Smith}, {Smith}, {Suchyta},
  {Swanson}, {Tarle}, {Thomas}, {Thomas}, {Troxel}, {Vikram}, {Wechsler},
  {Weller}, {Dark Energy Survey}, \& {Dark Energy Camera GW-EM
  Collaboration}}]{soaressantos17}
{Soares-Santos}, M., {Holz}, D.~E., {Annis}, J., {et~al.} 2017, \apjl, 848,
  L16, \dodoi{10.3847/2041-8213/aa9059}

\bibitem[{Sprouse {et~al.}(2024)Sprouse, Lund, Miller, McLaughlin, \&
  Mumpower}]{Sprouse_2024}
Sprouse, T.~M., Lund, K.~A., Miller, J.~M., McLaughlin, G.~C., \& Mumpower,
  M.~R. 2024, The Astrophysical Journal, 962, 79,
  \dodoi{10.3847/1538-4357/ad1819}

\bibitem[{Sugita {et~al.}(2018)Sugita, Kawai, Nakahira, Negoro, Serino, Mihara,
  Yamaoka, \& Nakajima}]{Sugita_2018}
Sugita, S., Kawai, N., Nakahira, S., {et~al.} 2018, Publications of the
  Astronomical Society of Japan, 70, \dodoi{10.1093/pasj/psy076}

\bibitem[{{Tanaka} {et~al.}(2017){Tanaka}, {Utsumi}, {Mazzali}, {Tominaga},
  {Yoshida}, {Sekiguchi}, {Morokuma}, {Motohara}, {Ohta}, {Kawabata}, {Abe},
  {Aoki}, {Asakura}, {Baar}, {Barway}, {Bond}, {Doi}, {Fujiyoshi}, {Furusawa},
  {Honda}, {Itoh}, {Kawabata}, {Kawai}, {Kim}, {Lee}, {Miyazaki}, {Morihana},
  {Nagashima}, {Nagayama}, {Nakaoka}, {Nakata}, {Ohsawa}, {Ohshima}, {Okita},
  {Saito}, {Sumi}, {Tajitsu}, {Takahashi}, {Takayama}, {Tamura}, {Tanaka},
  {Terai}, {Tristram}, {Yasuda}, \& {Zenko}}]{tanaka17}
{Tanaka}, M., {Utsumi}, Y., {Mazzali}, P.~A., {et~al.} 2017, PASJ,
  \dodoi{doi:10.1093/pasj/psx121}

\bibitem[{{Tanvir} {et~al.}(2017){Tanvir}, {Levan},
  {Gonz{\'a}lez-Fern{\'a}ndez}, {Korobkin}, {Mandel}, {Rosswog}, {Hjorth},
  {D'Avanzo}, {Fruchter}, {Fryer}, {Kangas}, {Milvang-Jensen}, {Rosetti},
  {Steeghs}, {Wollaeger}, {Cano}, {Copperwheat}, {Covino}, {D'Elia}, {de Ugarte
  Postigo}, {Evans}, {Even}, {Fairhurst}, {Figuera Jaimes}, {Fontes}, {Fujii},
  {Fynbo}, {Gompertz}, {Greiner}, {Hodosan}, {Irwin}, {Jakobsson},
  {J{\o}rgensen}, {Kann}, {Lyman}, {Malesani}, {McMahon}, {Melandri},
  {O'Brien}, {Osborne}, {Palazzi}, {Perley}, {Pian}, {Piranomonte}, {Rabus},
  {Rol}, {Rowlinson}, {Schulze}, {Sutton}, {Th{\"o}ne}, {Ulaczyk}, {Watson},
  {Wiersema}, \& {Wijers}}]{tanvir17}
{Tanvir}, N.~R., {Levan}, A.~J., {Gonz{\'a}lez-Fern{\'a}ndez}, C., {et~al.}
  2017, \apjl, 848, L27, \dodoi{10.3847/2041-8213/aa90b6}

\bibitem[{{Troja} {et~al.}(2017){Troja}, {Piro}, {van Eerten}, {Wollaeger},
  {Im}, {Fox}, {Butler}, {Cenko}, {Sakamoto}, {Fryer}, {Ricci}, {Lien}, {Ryan},
  {Korobkin}, {Lee}, {Burgess}, {Lee}, {Watson}, {Choi}, {Covino}, {D'Avanzo},
  {Fontes}, {Gonz{\'a}lez}, {Khandrika}, {Kim}, {Kim}, {Lee}, {Lee}, {Kutyrev},
  {Lim}, {S{\'a}nchez-Ram{\'{\i}}rez}, {Veilleux}, {Wieringa}, \&
  {Yoon}}]{troja17}
{Troja}, E., {Piro}, L., {van Eerten}, H., {et~al.} 2017, \nat, 551, 71,
  \dodoi{10.1038/nature24290}

\bibitem[{Utsumi {et~al.}(2017)Utsumi, Tanaka, Tominaga, Yoshida, Barway,
  Nagayama, Zenko, Aoki, Fujiyoshi, Furusawa, Kawabata, Koshida, Lee, Morokuma,
  Motohara, Nakata, Ohsawa, Ohta, Okita, Tajitsu, Tanaka, Terai, Yasuda, Abe,
  Asakura, Bond, Miyazaki, Sumi, Tristram, Honda, Itoh, Itoh, Kawabata,
  Morihana, Nagashima, Nakaoka, Ohshima, Takahashi, Takayama, Aoki, Baar, Doi,
  Finet, Kanda, Kawai, Kim, Kuroda, Liu, Matsubayashi, Murata, Nagai, Saito,
  Saito, Sako, Sekiguchi, Tamura, Tanaka, Uemura, \& Yamaguchi}]{Utsumi_2017}
Utsumi, Y., Tanaka, M., Tominaga, N., {et~al.} 2017, Publications of the
  Astronomical Society of Japan, 69, \dodoi{10.1093/pasj/psx118}

\bibitem[{{Valenti} {et~al.}(2017){Valenti}, {David}, {Sand}, {Yang},
  {Cappellaro}, {Tartaglia}, {Corsi}, {Jha}, {Reichart}, {Haislip}, \&
  {Kouprianov}}]{valenti17}
{Valenti}, S., {David}, {Sand}, J., {et~al.} 2017, \apjl, 848, L24,
  \dodoi{10.3847/2041-8213/aa8edf}

\bibitem[{{Wollaeger} \& {van Rossum}(2014)}]{wollaeger14}
{Wollaeger}, R.~T., \& {van Rossum}, D.~R. 2014, The Astrophysical Journal
  Supplement Series, 214, 28, \dodoi{10.1088/0067-0049/214/2/28}

\bibitem[{Wollaeger {et~al.}(2013)Wollaeger, van Rossum, Graziani, Couch,
  Jordan~IV, Lamb, \& Moses}]{wollaeger13}
Wollaeger, R.~T., van Rossum, D.~R., Graziani, C., {et~al.} 2013, The
  Astrophysical Journal Supplement Series, 209, 36,
  \dodoi{10.1088/0067-0049/209/2/36}

\bibitem[{{Wollaeger} {et~al.}(2018){Wollaeger}, {Korobkin}, {Fontes},
  {Rosswog}, {Even}, {Fryer}, {Sollerman}, {Hungerford}, {van Rossum}, \&
  {Wollaber}}]{wollaeger18}
{Wollaeger}, R.~T., {Korobkin}, O., {Fontes}, C.~J., {et~al.} 2018, \mnras,
  478, 3298, \dodoi{10.1093/mnras/sty1018}

\bibitem[{{Wollaeger} {et~al.}(2019){Wollaeger}, {Fryer}, {Fontes}, {Lippuner},
  {Vestrand}, {Mumpower}, {Korobkin}, {Hungerford}, \& {Even}}]{wollaeger19}
{Wollaeger}, R.~T., {Fryer}, C.~L., {Fontes}, C.~J., {et~al.} 2019, \apj, 880,
  22, \dodoi{10.3847/1538-4357/ab25f5}

\bibitem[{{Wollaeger} {et~al.}(2021){Wollaeger}, {Fryer}, {Chase}, {Fontes},
  {Ristic}, {Hungerford}, {Korobkin}, {O'Shaughnessy}, \&
  {Herring}}]{wollaeger21}
{Wollaeger}, R.~T., {Fryer}, C.~L., {Chase}, E.~A., {et~al.} 2021, \apj, 918,
  10, \dodoi{10.3847/1538-4357/ac0d03}

\end{thebibliography}

\end{document}